%% file: main.tex
\documentclass[sigconf]{acmart}
\usepackage{amsfonts}
\usepackage{amsmath}
\usepackage{bbm}
\usepackage{graphicx}
\usepackage{xspace}
\usepackage{mathtools,soul}
\usepackage[flushleft]{threeparttable}
\usepackage{multirow}
\usepackage{rotating}
\usepackage{booktabs, color}	
\usepackage{wrapfig, lipsum}
\usepackage{subcaption}
\usepackage{enumitem}
\usepackage{hyperref}
\usepackage{pgfplots}
\pgfplotsset{compat=1.17}
\usepackage{tikz}
\usepackage{scalerel}
\usetikzlibrary{shapes.geometric}

\graphicspath{ {./figures/} {./plots/} {./plots/sim/}  {./plots/runtimeLayers/} {./plots/oversmoothing/} {./plots/stability/} {./plots/stabilityDims/} {./plots/activity/} }
\newcommand{\etal}{\textit{et al}. }
\newcommand{\redline}{\raisebox{2pt}{\tikz{\draw[red, line width=1pt](0,0) -- (6mm,0); \fill[red] (3mm,0) circle (1.8pt);}}}
\newcommand{\blueline}{\raisebox{2pt}{\tikz{
\draw[blue,solid,line width=1pt](0,0) -- (6mm,0);
\node[star, star points=5, star point ratio=2.25, minimum size=5pt, fill=blue, inner sep=0pt] at (3mm, 0) {};
}}}
\newcommand{\magentaline}{\raisebox{2pt}{\tikz{
\draw[magenta,solid,line width=1pt](0,0) -- (6mm,0);
\node[rectangle, minimum size=5pt, fill=magenta] at (3mm, 0) {};
}}}

\input{define}

\AtBeginDocument{%
  \providecommand\BibTeX{{%
    \normalfont B\kern-0.5em{\scshape i\kern-0.25em b}\kern-0.8em\TeX}}}

\setcopyright{acmcopyright}
\copyrightyear{2024}
\acmYear{2024}
\acmDOI{XXXXXXX.XXXXXXX}

\acmConference[Conference acronym 'XX]{Make sure to enter the correct
  conference title from your rights confirmation emai}{June 03--05,
  2018}{Woodstock, NY}
\acmPrice{15.00}
\acmISBN{978-1-4503-XXXX-X/18/06}

\begin{document}

\title{Modeling Sequences as Star Graphs to Address
Over-smoothing in Self-attentive Sequential Recommendation}

\author{Bo Peng}
\affiliation{%
  \institution{The Ohio State University}
  \city{Columbus}
  \state{OH}
  \country{US}}
\email{peng.707@buckeyemail.osu.edu}

\author{Ziqi Chen}
\affiliation{%
  \institution{The Ohio State University}
  \city{Columbus}
  \state{OH}
  \country{US}}
\email{chen.8484@buckeyemail.osu.edu}

\author{Srinivasan Parthasarathy}
\affiliation{%
  \institution{The Ohio State University}
  \city{Columbus}
  \state{OH}
  \country{US}}
\email{srini@cse.ohio-state.edu}

\author{Xia Ning}
\affiliation{%
  \institution{The Ohio State University}
  \city{Columbus}
  \state{OH}
  \country{US}}
\email{ning.104@osu.edu}

\renewcommand{\shortauthors}{Trovato and Tobin, et al.}

\begin{abstract}
%Sequential recommendation aims to recommend the next item of users' interest based on their historical interactions.
%
%Recently, 
Self-attention (SA) mechanisms have been widely used in developing sequential recommendation (SR) methods, 
and demonstrated state-of-the-art performance. 
%
%Given a sequence, the self-attention mechanism calculates an attention weight for 
%each pair of items within the sequence, 
%and aggregate items using these attention weights.
%
%From a graph perspective, 
%this is equivalent to treating items within a sequence as nodes in a fully connected graph, where edge weights are determined by the attention weights.
%
%\bo{
%This design allows each item node to aggregate information from all others in one step.
%
%Thus, self-attention mechanisms could effectively capture long-range dependencies within sequences.
%}
%
%Compared to prevalent recurrent neural networks where items are aggregated sequentially, 
%
%by modeling sequences as fully connected graphs, 
%self-attention mechanisms allow each item node to aggregate information from all others in one step, and thus, could more effectively capture long-range dependencies within sequences.
%
%Though promising,
However, in this paper, 
we show that self-attentive SR methods substantially suffer from the over-smoothing issue that item embeddings within a sequence become increasingly similar across attention blocks.
%
%As shown in recent work, 
As widely demonstrated in the literature, 
this issue could lead to a loss of information in individual items, and significantly degrade models' scalability and performance.
%
%Learning similar embeddings for all items within a sequence leads to the loss of information on individual items, and thus, 
%could substantially deteriorate 
%the model expressiveness.
%
To address the over-smoothing issue, 
in this paper, 
%instead of a fully connected graph, 
we view items 
within a sequence constituting a star graph and develop a method, denoted as \method, for SR.
Different from existing self-attentive methods, 
\method introduces an additional internal node to specifically capture the global information within the sequence, and does not require information propagation among items. 
This design fundamentally addresses the over-smoothing issue 
and enables \method a linear time complexity with respect to the sequence length.
We compare \method with ten state-of-the-art
baseline methods on six public benchmark datasets. 
Our experimental results demonstrate that \method significantly outperforms the baseline methods, 
%on benchmark datasets
with an improvement of as much as 10.10\%. 
Our analysis shows the superior scalability of \method over 
the state-of-the-art 
self-attentive methods.
Our complexity analysis and run-time performance comparison together show that \method is both theoretically and practically more efficient than self-attentive methods.
Our analysis of the attention weights learned in SA-based methods indicates that on sparse recommendation data, modeling dependencies in all item pairs using the SA mechanism yields limited information gain, and thus, might not benefit the recommendation performance~\footnote{ 
We will release the source code and processed datasets upon the acceptance of this paper.
}.
\end{abstract}




\received{20 February 2007}
\received[revised]{12 March 2009}
\received[accepted]{5 June 2009}

\maketitle

%%%%%%%%%%%%%%%%%%%%%%%%%%%%%%%%%%%%%%%%%%%%%%%%%%%
\section{Introduction}
\label{sec:introduction}
%%%%%%%%%%%%%%%%%%%%%%%%%%%%%%%%%%%%%%%%%%%%%%%%%%%

\begin{figure*}[!h]
       \centering
       \footnotesize
        \begin{minipage}{\linewidth}
               \begin{subfigure}{0.33\linewidth}
                    \centering
                    \includegraphics[width=0.6\linewidth]{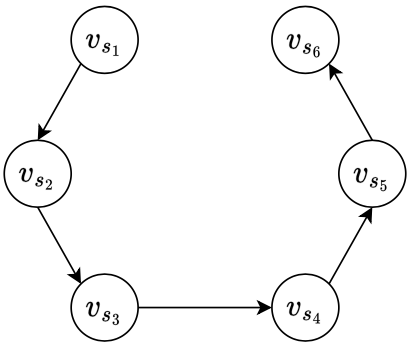}
                    \caption{Sequential Graph}
                    \label{fig:sequential_graph}
                \end{subfigure}
                \begin{subfigure}{0.33\linewidth}
                    \centering
                    \includegraphics[width=0.6\linewidth]{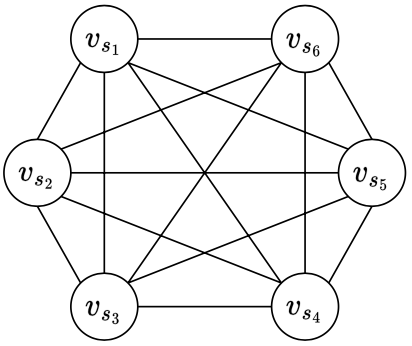}
                    \caption{Fully Connected Graph}
                    \label{fig:clique_graph}
                \end{subfigure}
                \begin{subfigure}{0.33\linewidth}
                    \centering
                    \includegraphics[width=0.6\linewidth]{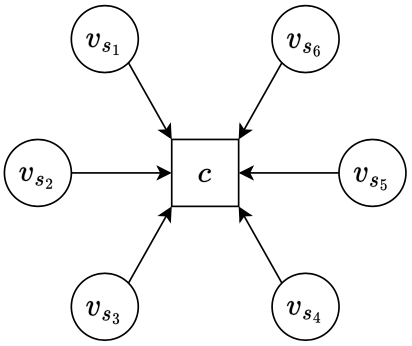}
                    \caption{Star Graph}
                    \label{fig:star_graph}
                \end{subfigure}
    \end{minipage}
\caption{
Illustration of a sequence represented by different graphs, 
in which $v_{s_t}$ is the $t$-th item in the sequence and $c$ is the internal node in the star graph.
%
%For simplicity, we drop the self-loop in the fully connected graph.
}
\label{fig:graph}
%https://app.diagrams.net/#G1afnO6u4IMe1FsFBPBSeklsOy4MUiY9Ue
\vspace{-10pt}
\end{figure*}

SR aims to 
identify and recommend the next item of users' interest based on their historical interactions.
It has been widely employed in applications such as online retail~\cite{he2016ups}
and video streaming~\cite{belletti2019quantifying}, 
and has been drawing increasing attention from the research community.
%
%A key challenge in developing modern SR methods could be to effectively aggregate information from interacted items and 
%capture the global 
%information within sequences.
%
With the prosperity of deep learning, 
deep neural networks, especially recurrent neural networks (RNNs)~\cite{hidasi2018recurrent} 
have emerged as the prominent architecture of SR methods.
From a graph perspective, 
RNNs represent each interaction sequence using a sequential graph as shown in Figure~\ref{fig:sequential_graph}, and recurrently integrate information from items within the sequence.
This design, however, may not 
effectively capture the long-range dependencies within sequences as 
it is challenging to propagate information across long distances~\cite{vaswani2017attention}.
%faces issues in 
%propagating information across long distances~\cite{vaswani2017attention}, and thus, may not effectively capture the long-range dependencies within sequences.
%

To better capture the long-range dependencies, SA mechanisms~\cite{vaswani2017attention} 
have recently been utilized for 
SR~\cite{sasrec,zhou2022filter}, 
and demonstrated state-of-the-art performance~\cite{sasrec}. 
Given a sequence, 
SA mechanisms learn attention weights for all pairs of items within sequences, 
and aggregates items using these attention weights.
From a graph perspective, this is essentially 
equivalent to viewing items 
within a sequence as nodes 
in a fully connected graph
~\footnote{For simplicity, 
we drop the self-loops in the graph.}
~\cite{dwivedi2020generalization} as illustrated in Figure~\ref{fig:clique_graph}, 
where $v_{s_t}$ is the $t$-th item in the sequence, and the edge weights are determined by the attention weights. 
%
%SA mechanisms differ from RNNs in that it essentially assumes items within a sequence constitute a complete graph (Figure~\ref{fig:clique_graph}) in which the edge weights are determined by the attention weights~\cite{shi2022revisiting}. 
%
The fully connected graph allows 
each item node to propagate information 
to all others in a single step, 
thereby enabling SA-based methods to effectively capture long-range dependencies within a sequence.
%
%Consequently, in SA mechanisms, 
%each item node could propagate information 
%to all others in a single step, 
%thereby enabling SA to effectively capture long-range dependencies within a sequence.
%

Though promising, 
%\bo{
modeling sequences as fully connected graphs 
leads to two main issues.
%}
%allowing each item node to integrate information 
%from all others could lead to two main issues.
%
First, 
as will be demonstrated in our analysis (Section~\ref{sec:results:smoothing}), 
SA-based SR methods substantially suffer from the 
over-smoothing issue~\cite{chen2020measuring} 
that the embeddings of items within a sequence could become more and more similar across attention blocks.
Learning similar embeddings for all items within a sequence leads to the loss of information on individual items, and thus, 
could substantially degrade the model performance~\cite{chen2020measuring,shi2022revisiting,huang2020tackling,xu2018representation}.
In addition, as shown in Shi~\etal~\cite{shi2022revisiting}, the over-smoothing issue also deteriorates the learning of deep models, thereby degrading the model scalability.
Second, 
SA-based SR methods generally suffer from quadratic time complexities, which limit their utilities in large-scale recommendation applications.

Addressing the above issues while still effectively 
capturing the long-range dependencies within 
sequences represents a critical research challenge 
in developing SR methods.
To tackle this challenge, 
in this paper, we develop \method, 
a method that models user interaction sequences using star graphs for SR.
Specifically,
\method stacks multiple blocks 
to recursively capture users' intent 
from their interaction sequence~\cite{sasrec}.
\method has an attention 
layer and a feed-forward layer in each block.
In the attention layer, 
as illustrated in Figure~\ref{fig:star_graph}, 
\method models each sequence as a star graph, where edge weights are determined by the attention weights,  
and introduces an additional internal node $c$ to integrate information from all item nodes $v_{s_t}$, thereby capturing the user's intent.
In the feed-forward layer, 
\method updates the embedding 
of the internal node via non-linear 
networks to more accurately estimate 
the user's intent.
Previous work~\cite{chen2020measuring} 
shows that requiring information propagation among all item nodes could be the essential reason for the over-smoothing issue.
Thus, \method does not propagate information among items. 
Instead, it utilizes an internal node for information aggregation.
This design allows \method to fundamentally address the over-smoothing issue (Section~\ref{sec:results:smoothing}).
Beyond addressing the 
over-smoothing issue, 
by modeling sequences as star graphs, 
\method also achieves a linear time complexity with respect to the sequence length as will be proven in Section~\ref{sec:method:complexity}.
In addition,  
similarly to that in SA mechanisms, 
\method could effectively capture long-range dependencies within sequences as each item node could propagate information to the internal 
node within one step.

We extensively compare \method with ten state-of-the-art baseline methods on six benchmark datasets.
Our experimental results demonstrate that \method could remarkably outperform 
the state-of-the-art baseline methods 
on the six datasets, 
with an improvement of up to 10.10\% at Recall@$10$ (Section~\ref{sec:materials:protocl:metric}). 
Our experimental results also show that 
\method consistently outperforms 
the state-of-the-art SA-based SR method on 
users with different activity levels (Section~\ref{sec:results:ufreq}).
In addition, 
our analysis reveals that existing 
SA-based SR methods suffer from the over-smoothing issue (Section~\ref{sec:results:smoothing}), 
which limits their scalability in learning 
deep models (Section~\ref{sec:results:scalability}).
Our complexity analysis (Section~\ref{sec:method:complexity}) and run-time performance comparison (Section~\ref{sec:results:runtime}) 
together show that 
\method is both theoretically and practically more efficient than the state-of-the-art
SA-based SR methods.
%
%\bo{
Our analysis also suggests that on sparse recommendation data, leveraging SA mechanisms to capture item dependencies could yield limited information gain.
%}

We summarize our major contributions as follows: 
1) To the best of our knowledge, this is the first work 
demonstrating the over-smoothing issue in SA-based SR methods; 
2) We develop \method for SR in which we model sequences using star graphs to address 
the over-smoothing issue and enable linear time complexity; 
3) %Our experimental results show that \method outperforms ten state-of-the-art baseline methods on six benchmark datasets;
\method outperforms ten state-of-the-art baseline methods on six benchmark datasets;
4) Our analysis suggests that \method could achieve both superior scalability and superior run-time performance 
compared to SA-based SR methods;
%5) Our complexity analysis proves that \method is theoretically more efficient than SA-based methods, 
%and our run-time performance comparison 
%demonstrates that \method is also 
%practically more efficient than SA-based methods on GPUs.
%\bo{
5) Our analysis also indicates that 
modeling item dependencies using the SA mechanism on sparse recommendation data could yield limited information gain, and thus, might not improve the recommendation performance. 
%}
%6) We exhaustively tune the hyper-parameters of \method and baseline methods to enable a fair comparison.

%%%%%%%%%%%%%%%%%%%%%%%%%%%%%%%%%%%%%%%%%%%%%%%%%%%
\section{Related Work}
\label{sec:literature}
%%%%%%%%%%%%%%%%%%%%%%%%%%%%%%%%%%%%%%%%%%%%%%%%%%%

%**************************************************
\subsection{Sequential Recommendation}
\label{sec:literature:sequential}
%**************************************************

%Sequential recommendation aims to recommend the next item of users' interest 
%based on their historical interactions.
%
Numerous SR methods have been developed, 
particularly using Markov Chains (MCs) 
and neural networks.
Specifically, Rendle~\etal~\cite{rendle2010factorizing} developed \FPMC, a method 
in which MCs are utilized to model the transitions among items.
In recent years, neural networks 
such as RNNs 
and convolutional neural networks (CNNs)  
have been widely adapted for SR.
For instance, 
%Hidasi~\etal~\cite{hidasi2018recurrent}
%~\cite{hidasi2015session} 
%developed a gated recurrent units (GRUs) based method \GRURec 
%in which GRUs are employed to recurrently model users' preferences.
%
%Hidasi~\etal also extended \GRURec to \GRURecP~\cite{hidasi2018recurrent} that
%uses a novel ranking loss to mitigate the 
%degradation problem in GRUs when processing long sequences.
%
Li~\etal~\cite{li2017neural} developed \NARM, which 
incorporates attention mechanisms into RNNs 
to better capture users' long-term preferences.
%Vasile~\etal~\cite{vasile2016meta} developed a skip-gram-based method 
%to leverage the skip-gram model~\cite{mikolov2013distributed} to capture the co-occurrence among items for recommendation.
%
%Recently, CNNs are also adapted 
%for sequential recommendation.
%
Tang~\etal~\cite{tang2018personalized} developed a CNNs-based model \Caser that employs vertical and horizontal convolutional filters to capture 
the synergies among items for better recommendation.
%
%Yuan~\etal~\cite{yuan2018simple} developed a CNNs-based generative model \NextItRec 
%that could better capture users' long-term preferences compared to \Caser.
%
Besides RNNs and CNNs-based methods, 
SA-based methods are also widely developed 
for SR.
Kang~\etal~\cite{sasrec} developed a SA-based method \SASRec, 
which stacks multiple SA blocks 
to recursively learn users' preferences.
Sun~\etal~\cite{sun2019bert4rec} integrated the cloze objective and SA mechanisms, and developed a bidirectional sequence modeling method, denoted as \BERT, for comprehensive user intent modeling.
Zhou~\etal~\cite{zhou2022filter} developed \FMLP for SR, in which they replaced the attention map within each SA block with a predefined transformation matrix for more efficient item aggregation.
He~\etal~\cite{he2021locker} constrained SA blocks to attend to local items, thereby enhancing the modeling of users' short-term preferences.
Recently, simple shallow methods 
have been introduced for SR, and shown promising performance. 
For example, Ma~\etal~\cite{ma2019hierarchical} developed \HGN, 
which uses a single gating layer to adaptively aggregate items and capture users' preferences.
Peng~\etal~\cite{peng2021ham} developed \HAM 
in which a single pooling layer is 
employed to learn the 
associations among items.

We notice that 
\SGNN developed in Pan~\etal~\cite{pan2020star} also leverages star graphs for SR.
However, \method differs significantly from \SGNN.
Specifically, \method is developed to address the over-smoothing issue in SA-based SR methods.
To this end, \method utilizes the internal node to aggregate information from individual items, 
and does not allow information propagation among items.
In contrast, \SGNN aims to improve graph neural networks (GNNs) in capturing long-range dependencies in sequences, and leverages the internal 
node as an anchor node to better propagate information among items. 
\method and \SGNN are developed to address 
different issues, 
and have substantially different architectures.
Thus, \method is not an extension of \SGNN.

%We noticed that \SGNN developed in Pan~\etal~\cite{pan2020star} also leveraged star graphs for SR.
%
%However, \SGNN differs significantly from \method.
%
%The key distinction between \method and \SGNN is that 
%\method does not allow information propagation among items to address the over-smoothing issue. 
%
%In contrast, \SGNN constructed a bidirectional star graph and just leveraged the internal node ($c$ in Figure~\ref{fig:star_graph}) as an anchor node to propagate information among items.
%
%Consequently, different from \method, \SGNN cannot address the over-smoothing issue.
%

\vspace{-10pt}

%************************************************** 
\subsection{Over-smoothing}
\label{sec:literature:attention}
%**************************************************

The over-smoothing issue was firstly 
identified in GNNs~\cite{hamilton2017inductive} by Li~\etal~\cite{li2018deeper}.
They observed that as GNNs propagate and mix information 
between neighboring nodes across layers, 
all nodes could eventually have identical embeddings.
As a result, GNNs lose the information on individual nodes, 
which could substantially deteriorate the model expressiveness~\cite{chen2020measuring}.
Recently, 
Shi~\etal~\cite{shi2022revisiting} has proven that theoretically, 
SA-based methods could also suffer from over-smoothing and 
result in sub-optimal performance.
Numerous approaches have been introduced 
to mitigate this issue.
For example, 
Chen~\etal~\cite{chen2020measuring} 
optimized the graph topology based on the model prediction to mitigate over-smoothing in GNNs.
%
%Zhao~\etal~\cite{zhao2019pairnorm} introduced a parinorm layer to prevent all node embeddings from becoming too similar.
%
%\bo{
Shi~\etal~\cite{shi2022revisiting} developed  hierarchical fusion strategies (\Concat and \Max) to alleviate over-smoothing by fusing embeddings from both 
shallow layers and deep layers as 
final output. 
%}
%Recently, Shi~\etal~\cite{shi2022revisiting} developed a hierarchical fusion strategy that adaptively combines the output from different layers to mitigate the over-smoothing issue.
%
%However, 
%to the best of our knowledge, none of the existing approaches was developed for SR.

%%%%%%%%%%%%%%%%%%%%%%%%%%%%%%%%%%%%%%%%%%%%%%%%%%%
\section{Definition and notations}
\label{sec:notation}
%%%%%%%%%%%%%%%%%%%%%%%%%%%%%%%%%%%%%%%%%%%%%%%%%%%

%\input{tables/notations.tex}

In this paper, we tackle the SR problem that given the historical interactions of users, we recommend the next item of users' interest.
In this paper, $\mathbb{U} = \{u_1, u_2, \dots\}$ is the set of all the users, where $u_i$ represents the $i$-th user, and $|\mathbb{U}|$ is the total number of users.
Similarly, we represent the set of all the items as $\mathbb{V} = \{v_1, v_2, \dots\}$. 
$|\mathbb{V}|$ is the total number of items.
The historical interactions of $u_i$ is represented as a sequence $S_i = \{v_{s_1}(i), v_{s_2}(i), \dots\}$, where $v_{s_t}(i)$ is the $t$-th interacted item in $S_i$ and $|S_i|$ is the length of the sequence.
Given $S_i$, 
the next item that $u_i$ will interact with 
is referred to as the ground-truth next item, denoted as $v_{g}(i)$.
The goal of \method is to correctly recommend $v_{g}(i)$ for $u_i$.
When no ambiguity arises, we will eliminate $i$ in $S_i$, $v_{s_t}(i)$ and $v_{g}(i)$. 
We use uppercase letters to denote matrices, lower-case bold letters to
denote row vectors and lower-case non-bold letters to represent scalars.
Table~\ref{tbl:notations} (Appendix~\ref{sec:not}) shows the key notations used in this paper.

\vspace{0.5em}
\noindent
\textit{\textbf{Over-smoothing in SA-based methods}: 
the embeddings of items within a sequence 
become increasingly similar across attention blocks, thereby deteriorating the model scalability and performance.}

%%%%%%%%%%%%%%%%%%%%%%%%%%%%%%%%%%%%%%%%%%%%%%%%%%%
\section{Method}
\label{sec:method}
%%%%%%%%%%%%%%%%%%%%%%%%%%%%%%%%%%%%%%%%%%%%%%%%%%%

%\bo{Add a figure for the architecture}

\begin{figure}[t]
	%\vspace{-10pt}
	\includegraphics[width=\linewidth]{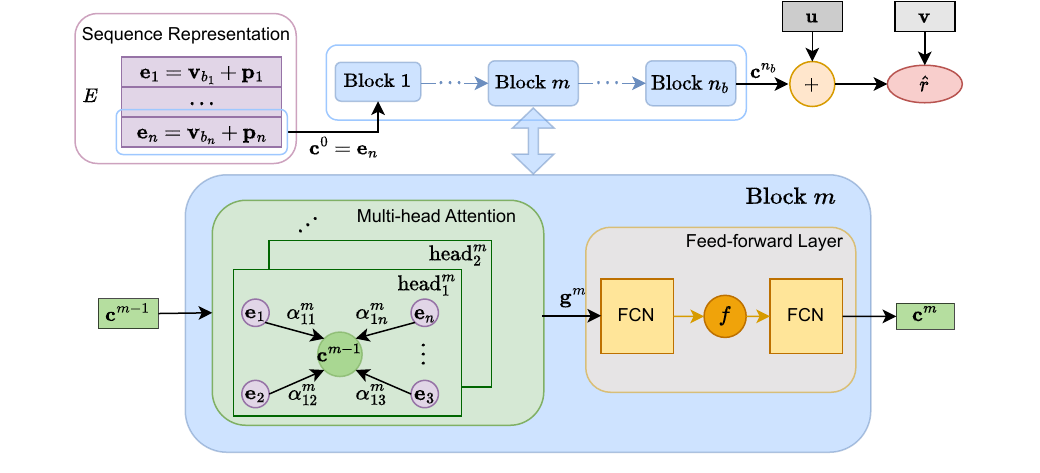}
    %https://app.diagrams.net/#G1-9hb23OcGyWUZeh1efQ8vGR06qntU85F
    %https://app.diagrams.net/#G187xLlKr0L7UTqM4Goy3y6x6gQ7_CxrWt
	\vspace{-10pt}
	\caption{
    The overall architecture of \method.
    \method models sequences using star graphs and utilizes the internal node $c$ to integrate information from all item nodes.}
	\label{fig:architecture}
	\vspace{-10pt}
\end{figure}

Figure~\ref{fig:architecture} presents the overall architecture of \method. 
\method treats sequences as star graphs to address the over-smoothing issue. 
\method also explicitly captures users' long-term preferences to enable comprehensive user intent modeling.
We present the detail of \method in the following sections.

%**************************************************
\subsection{Sequence Representation}
\label{sec:method:seq}
%**************************************************

Following \SASRec, we focus on the most recent $n$ items in users' interaction histories to generate recommendations.
Specifically, we transform each interaction sequence $S$ to a fixed-length sequence $B$ = $\{v_{b_1}, v_{b_2} \dots\}$, which contains the last $n$ items in $S$ (i.e., $v_{b_t}$=$v_{s_{|S_i|-n+t}}$).
If $|S|$ is shorter than $n$, we pad empty items, denoted as $v_0$, 
at the beginning of $B$.
In \method, following the literature~\cite{sasrec,vaswani2017attention}, we represent items and positions in each sequence using learnable embeddings.
We learn an item embedding matrix $V \in \mathbb{R}^{|\mathbb{V}| \times d}$ in which the $j$-th row $\mathbf{v}_j$ 
is the embedding of item $v_j$,  
%$|\mathbb{V}|$ is the total number of items in the dataset, 
and $d$ is the dimension of embeddings.
We use a constant zero vector as the embedding of padding items.
Similarly, we learn a position embedding matrix $P \in \mathbb{R}^{n \times d}$, and $\mathbf{p}_t$ is the embedding of the $t$-th position.
Given $V$ and $P$, we represent the sequence $B$ using a matrix $E$:
\begin{equation}
	\label{eqn:seq}
	E = [\mathbf{e}_1; \mathbf{e}_2; \cdots; \mathbf{e}_n] =  [\mathbf{v}_{b_1}+\mathbf{p}_{1}; \mathbf{v}_{b_2}+\mathbf{p}_{2}; \cdots; \mathbf{v}_{b_n}+\mathbf{p}_{n}],
\end{equation}
where $\mathbf{v}_{b_t}$ is the embedding of $v_{b_t}$.

%**************************************************
\subsection{Modeling Sequences as Star Graphs}
\label{sec:method:attention}
%**************************************************

%
\method employs a multi-block architecture 
to recursively model 
users' intent.
Distinct from existing SA-based SR methods~\cite{sasrec,fan2022sequential},
in each block, 
\method treats each item within $B$ 
as a node in a star graph, 
and introduces an additional internal node to 
estimate the user's intent by aggregating information from all item nodes using multi-head attention~\cite{vaswani2017attention}.
Across blocks, 
\method updates only the embedding of the internal node to enhance the estimation of the user's intent, 
%more accurately estimate the user's intent, 
while maintaining fixed embeddings on item nodes to 
preserve their individual information. 
In \method, we have $n_b$ blocks, 
and each block contains an attention layer  
and a feed-forward layer.
In what follows, we present the 
attention layer and the feed-forward layer in each block in detail.

%**************************************************
\subsubsection{Attention layer}
\label{sec:method:attention:attention}
%**************************************************

\method employs multi-head attention to adaptively aggregate information from item nodes  
to the central internal node in each block as follows:
\begin{eqnarray}
	\label{eqn:multi}
	\begin{aligned}
		\boldsymbol{\alpha}^{m}_k &= \text{softmax} \left (\frac{\left (\mathbf{c}^{m-1} Q^{m}_k \right ) 
        \left (E Z^{m}_k \right)^\top}{\sqrt{d}} \right ) , 
        \quad
		\mathbf{head}^{m}_k = \boldsymbol{\alpha}^{m}_k (E A^{m}_k), \\
		\mathbf{g}^{m} &= [\mathbf{head}^{m}_1, \mathbf{head}^{m}_2, \dots, \mathbf{head}^{m}_{n_h}] O^{m},\\
	\end{aligned}
\end{eqnarray}
where $\mathbf{c}^{m-1}$ is the embedding of the internal node from the $(m\text{-}1)$-th block, and 
$\mathbf{g}^{m}$ is the output of the attention layer in the $m$-th block.
Each attention layer has $n_h$ attention heads and $\mathbf{head}^{m}_k$ is the output from the $k$-th attention head. 
$\boldsymbol{\alpha}^{m}_k$ is the attention weights learned in the $k$-th attention head. 
%and $\alpha^{m}_{kt}$ is the attention weight on the $t$-th item in $B$.
%
$Q^{m}_k, Z^{m}_k \text{and } A^{m}_k \in \mathbb{R}^{d \times \frac{d}{n_h}}$ are learnable parameters in the $k$-th attention head.
$O^{m} \in \mathbb{R}^{d \times d}$ is a learnable parameter to integrate the attention heads.

Unlike SA-based methods, which require every item to aggregate information from all others to capture global information within the sequence, 
\method introduces an additional internal node to specifically capture the global information and estimate the user's intent. 
This approach results in a linear time complexity as will be shown in Section~\ref{sec:method:complexity}, while could still effectively capture long-range dependencies within the sequence.

%**************************************************
\subsubsection{Feed-forward layer}
\label{sec:method:attention:feed} 
%**************************************************

After each attention layer, 
we include a feed-forward layer
to endow \method with non-linearity, 
and enable more expressive models~\cite{vaswani2017attention}.
Particularly, given $\mathbf{g}^{m}$, 
we stack two fully-connected networks (FCNs) as the feed-forward layer:
\begin{equation}
	\label{eqn:feed}
	\mathbf{c}^{m} = (f(\mathbf{g}^{m}W^{m}_1+\mathbf{b}^{m}_1))W^{m}_2 + \mathbf{b}^{m}_2,
\end{equation}
where $f(\cdot)$ is an activation function such as ReLU~\cite{nair2010rectified} and GELU~\cite{hendrycks2016gaussian}, 
$W^{m}_1$, $W^{m}_2 \in \mathbb{R}^{d \times d}$ and $\mathbf{b}^{m}_1$, $\mathbf{b}^{m}_2$ are learnable parameters.
It is worth noting that different from SA-based methods,
\method updates only the embedding of the internal node across blocks, 
while fixing the item node embeddings.
As a result, 
\method could preserve the information on individual items and will not be affected by the over-smoothing issue, even when learning deep models (Section~\ref{sec:results:smoothing}).
Following \SASRec~\cite{sasrec}, 
we connect each layer with its previous layer using residual connections~\cite{he2016deep}
to mitigate the degradation issue~\cite{roy2018effects} 
during training. 
%and implicitly integrate outputs 
%from all layers
%to generate recommendations.
%
Recent studies~\cite{peng2021ham,stamp} show that the most recent interacted item could be a strong indicator of the next item of users' interest.
Thus, for each sequence, 
\method employs $\mathbf{e}_n$, 
the embedding of the last item in $E$ (Equation~\ref{eqn:seq}), 
as the initial embedding of 
the internal node in the first block 
(i.e., $\mathbf{c}^{0}$=$\mathbf{e}_n$).
%for better user intent modeling.

%**************************************************
\subsection{Comprehensive User Intent Modeling}
\label{sec:method:long}
%**************************************************

As demonstrated in the literature~\cite{peng2021ham,ma2019hierarchical}, both users' 
short-term preferences and long-term preferences play important roles in generating accurate recommendations. 
By considering $\mathbf{e}_n$, 
the embedding of the most recent interacted item, 
as the initial embedding of the internal node, 
the internal node embedding should be able to
effectively capture the user's short-term preferences across blocks.
However, it may not also fully capture the user's 
long-term preferences as the long-term preferences of the user could be different from her short-term preferences. 
%\method may not fully capture users' long-term preferences.
%
Thus, to enable comprehensive user intent modeling, \method learns user embeddings to specifically capture users' long-term preferences. 
In particular, \method learns an embedding matrix $U \in \mathbb{R}^{|\mathbb{U}| \times d}$ to represent the long-term preferences of all the users. 
The $i$-th row $\mathbf{u}_i$ in $U$ 
represents the long-term preferences of user $u_i$.
%

%**************************************************
\subsection{Recommendation Scores and Network Training}
\label{sec:method:scores}
%**************************************************

%Given the output of the attention blocks and the user embeddings, 
%we generate recommendation scores for candidate items as follows:
To generate recommendation scores for user $u_i$, \method integrates the embedding of the internal node from the last block (i.e., $\mathbf{c}_i^{n_b}$) 
and the learned user embedding $\mathbf{u}_i$, as follows:
\begin{equation}
	\label{eqn:score}
	\hat{r}_{ij} = (\mathbf{c}^{n_b}_i + \mathbf{u}_i) \mathbf{v}_j^\top, 
\end{equation}
where $\hat{r}_{ij}$ is the recommendation score of user $u_i$ on item $v_j$, 
and $n_b$ is the total number of blocks in \method.  
For each user, we recommend items with the top-$k$ highest recommendation scores.
%
%When no ambiguity arises, we will eliminate $i$ in $\hat{r}_{ij}$. 
%
Following \SASRec, 
we employ the binary cross-entropy loss
to minimize the negative log-likelihood of correctly recommending the ground truth next item as follows:
\begin{equation}
	\label{eqn:obj:primary}
	\min\limits_{\boldsymbol{\Theta}} -\sum \nolimits_{S_i \in \mathbb{T}, v_j \not \in S_i} \left[ \log(\sigma(\hat{r}_{ig})) + \log(1-\sigma(\hat{r}_{ij})) \right],
\end{equation}
%\xia{the loss is not correct! fix it}
where $\mathbb{T}$ is the set of all the training sequences; 
$\boldsymbol{\Theta}$ is the set of learnable parameters (e.g., $V$ and $P$); and  
$\hat{r}_{ig}$ is the recommendation 
score of user $u_i$ on her ground-truth next item
$v_{g}(i)$.
For each training sequence, we randomly sample one negative item $v_j$ for optimization.
All the learnable parameters are randomly initialized, 
and are optimized in an end-to-end manner.

%**************************************************
\subsection{Complexity Analysis}
\label{sec:method:complexity}
%**************************************************

%
Previous work~\cite{tay2020efficient,wang2020linformer,katharopoulos2020transformers} shows that 
the time complexity of each SA block is $\mathcal{O} (n^2d+nd^2)$, where $n$ is the length of the transformed sequence (Equation~\ref{eqn:seq}) 
and $d$ is the dimension of embeddings. 
The quadratic time complexity limits the utility of SA-based methods in latency-sensitive recommendation applications.
In contrast, 
\method models sequences as star graphs and could achieve 
a time complexity of $\mathcal{O} (nd^2)$, 
which is linear with respect to the sequence length $n$. 
This allows \method to be theoretically 
more efficient than SA-based methods.
Particularly, each SA block requires $6nd^2+2n^2d+2nd$ operations to compute, 
whereas each block in \method requires only $2nd^2+4d^2+2nd+2d$ operations.
The difference amounts to $4d^2(n-1) + 2d(n^2-1)$, 
which is quadratic with respect to both $d$ and $n$.
As will be shown in Section~\ref{sec:results:runtime}, 
the better time complexity could translate into 
superior run-time performance of \method 
over SA-based methods on modern GPUs.  

%%%%%%%%%%%%%%%%%%%%%%%%%%%%%%%%%%%%%%%%%%%%%%%%%%%
\section{Materials}
\label{sec:materials}
%%%%%%%%%%%%%%%%%%%%%%%%%%%%%%%%%%%%%%%%%%%%%%%%%%%

\subsection{Baseline Methods}
\label{sec:materials:baseline}

We compare \method with ten state-of-the-art baseline methods. 
Specifically, we compare \method with the  
MC-based method \FPMC~\cite{rendle2010factorizing} 
and the CNNs-based method \Caser~\cite{tang2018personalized}.
We also compare \method with the RNNs-based method \NARM~\cite{li2017neural} 
and two shallow methods \HAM~\cite{peng2021ham} and \HGN~\cite{ma2019hierarchical}.
We compare \method with three state-of-the-art SA-based methods \SASRec~\cite{sasrec}, \BERT~\cite{sun2019bert4rec} and \FMLP~\cite{zhou2022filter}.
%
%\bo{
We further compare \method with 
\SASRec equipped with two hierarchical fusion strategies: 1) \SASRec with \Concat (\SC); 
and 2) \SASRec with \Max (\SM).
These two fusion strategies are developed in Shi~\etal~\cite{shi2022revisiting} to alleviate over-smoothing. %in SA-based methods.
%
%}
%
%We refer the audience to Section~\ref{sec:literature} for a detailed description of these baseline methods.
%
Note that, we do not consider baseline methods that the implementation is not publicly available (e.g., \SGNN) to enable a fair comparison.
\HGN and \SASRec have been compared with 
a comprehensive
set of other methods including \GRURec~\cite{hidasi2018recurrent} and \NextItRec~\cite{yuan2018simple}, 
and have outperformed those methods. 
Therefore, we compare \method with \HGN and \SASRec instead of
the methods that they outperform.
%
%For better reproducibility, we report the implementation detail of \method and baseline methods in the appendix (Section~\ref{sec:reproducibility}).

%\bo{
To ensure a fair comparison, we exhaustively tune the hyper-parameters of all the baseline methods.
Thus, the performance of baseline methods reported in our experiments is generally better than that reported in the literature~\cite{ma2019hierarchical,rajput2023recommender}.
For example, on the \Beauty dataset (Section~\ref{sec:materials:datasets}), the Recall@10 (Section~\ref{sec:materials:protocl:metric}) of \SASRec reported in our experiments is 20.3\% higher than that reported in a recent work~\cite{rajput2023recommender}.
We report the implementation detail of \method and baseline methods, and the search range for each hyper-parameter in Appendix~\ref{sec:reproducibility}. 
%}

%**************************************************
\subsection{Datasets}
\label{sec:materials:datasets}
%**************************************************

%\bo{
We evaluate \method and baseline methods 
on six public benchmark datasets:
1) Amazon-Beauty (\Beauty) and Amazon-Toys (\Toys) 
are from Amazon reviews~\cite{amazon}; %~\footnote{https://jmcauley.ucsd.edu/data/amazon/};
2) Goodreads-Children (\Children) and Goodreads-Comics
(\Comics) are from the Goodreads website~\cite{goodreads1,goodreads2};
%~\footnote{https://sites.google.com/eng.ucsd.edu/ucsdbookgraph/home}; 
and
3) MovieLens-1M (\MLOM) and MovieLens-20M (\MLTM) are from the MovieLens website~\cite{movielens},
%~\footnote{https://movielens.org/}.
%
We discuss the dataset pre-processing and statistics in Appendix~\ref{sec:datasets_statistics}.
%}

%**************************************************
\subsection{Experimental Protocol}
\label{sec:materials:protocl}
%**************************************************

\subsubsection{Training, validation and testing sets}
\label{sec:materials:protocl:training}
 
Following \SASRec, on all the datasets, 
given the historical interactions $S$ of each user, 
we use the last item (i.e., $v_{s_{|S|}}$) in the history for testing, 
the second last item (i.e., $v_{s_{|S|-1}}$) for validation, 
and all the previous items for training. 
%
%Also following \SASRec, 
%for each training sequence $S$, we extract 
%the last $n$ items $\{b_1, b_2, \dots, b_n\}$, 
%and split it to $\{b_1, b_2\}$, $\{b_1, b_2, b_3\}$, 
%$\cdots$, $\{b_1, b_2, \dots, b_n\}$.
%
%We use all the resulted sequences 
%as the augmented sequences 
%(with necessary padding) for training.
%
 %
We tune hyper-parameters on the validation sets 
for all the methods using grid search, 
and use the best-performing hyper-parameters in terms of Recall@$10$ (Section~\ref{sec:materials:protocl:metric}) for testing.
%

%(Section~\ref{sec:reproducibility}).

%--------------------------------------------------
\subsubsection{Evaluation metrics}
\label{sec:materials:protocl:metric}
%--------------------------------------------------

Following the literature~\cite{sasrec,peng2021ham,ma2019hierarchical,tang2018personalized}, 
we use Recall@$k$ and NDCG@$k$ to evaluate \method and the baseline methods.
We refer the audience of interest to Peng~\etal~\cite{peng2021ham} for the detailed definitions of both Recall@$k$ and NDCG@$k$.
%\begin{itemize}
%Recall@$k$ measures the proportion of sequences 
%in which the ground-truth next item (i.e., $v_{s_{|S|-1}}$ during validation and $v_{s_{|S|}}$ during testing) is correctly recommended.
%
%Particularly, for each sequence, the Recall@$k$ is $1$ 
%if the ground-truth next item is at the top-$k$ of the recommendation list, 
%or $0$ otherwise.
%
%NDCG@$k$ is the normalized discounted cumulative gain. 
%It is a widely used rank-aware metric to evaluate 
%the ranking quality of recommendation lists.
%
%Following the literature~\cite{sasrec,peng2021ham}, 
%in our experiments, the gain indicates whether the ground-truth next item is recommended (i.e., gain is $1$) or not (i.e., gain is $0$).	
%\end{itemize}
% 
%In our experiments, 
%we report the average results over all the users 
%for each evaluation metric.
% 
%We also assess the statistical significance of performance differences at these evaluation metrics using the paired 
%t-test.
%

%%%%%%%%%%%%%%%%%%%%%%%%%%%%%%%%%%%%%%%%%%%%%%%%%%%
\section{Experimental Results}
\label{sec:results}
%%%%%%%%%%%%%%%%%%%%%%%%%%%%%%%%%%%%%%%%%%%%%%%%%%%

%**************************************************
\subsection{Overall Performance}
\label{sec:results:overall}
%**************************************************

\input{tables/overall}
%\label{tbl:overall_performance}

Table~\ref{tbl:overall_performance} presents the overall performance of \method, 
its variant \methodmu
and the state-of-the-art baseline methods at Recall@$k$ and NDCG@$k$ on six benchmark datasets.
In \methodmu, 
we remove user embeddings (i.e., $\mathbf{u}_i$) when calculating recommendation scores (Equation~\ref{eqn:score}).
We introduce \methodmu to evaluate the effectiveness of 
the learnable user embeddings in \method.
For \NARM, on \MLTM, with the implementation provided by the authors, 
we get the out-of-memory (OOM) issue 
on NVIDIA Volta V100 GPUs with 16 GB memory.

As shown in Table~\ref{tbl:overall_performance}, overall, 
\method is the best-performing method on the six datasets.
In terms of Recall@$10$ and Recall@$20$, \method significantly outperforms all the baseline methods on five out of the six datasets except \MLTM.
%
%Similarly, at Recall@$20$, \method also significantly outperforms all the baseline methods on four out of the six datasets (i.e., \Beauty, \Toys, \Children and \Comics), 
%and achieves highly competitive performance with the best-performing baseline method on \MLOM.
% 
We observe a similar trend at both NDCG@$10$ and NDCG@$20$.
%
%In terms of NDCG@$10$ and NDCG@$20$, \method achieves the best performance on five out of the six datasets except for \MLTM.
%
We notice that on \MLTM, 
\method considerably underperforms \SASRec.
%the performance of \method is considerably worse than that of \SASRec.
%the best-performing baseline method \SASRec.
%
However, 
without user embeddings, \methodmu could still 
significantly outperform \SASRec on \MLTM.
These results demonstrate the strong performance of \method and its variant \methodmu
on different recommendation scenarios.

%--------------------------------------------------
\subsubsection{Comparing \method to \FPMC and \Caser}
\label{sec:results:overall:fpmc}
%--------------------------------------------------

As presented in Table~\ref{tbl:overall_performance}, 
\method outperforms the MC-based method 
\FPMC and CNNs-based method \Caser 
by a significant margin 
on all the six benchmark datasets.
%
%Particularly, \method outperforms \FPMC and \Caser by a significant margin on all the six datasets.
%
%Similarly, \method also achieves superior performance over \Caser on \Beauty, \Toys, \Children and \Comics.
%
%On \MLOM and \MLTM, \method still achieves comparable performance compared to \Caser.
%
\FPMC views the transitions among items as an MC and leverages only the most recent interacted items to predict the next item of users' interest.
Consequently, \FPMC may not be able to fully utilize the information in early interacted items, and result in sub-optimal performance.
\Caser employs CNNs to integrate items within sequences.
Though CNNs have been shown effective in capturing local structures, they may struggle to capture global information within sequences~\cite{yamashita2018convolutional}.
In contrast, by modeling sequences as star graphs, 
\method could effectively aggregate information from all items to capture the global information, 
and thus, enable superior recommendation performance over \FPMC and \Caser.
%

%--------------------------------------------------
\subsubsection{Comparison between \method and \NARM}
\label{sec:results:overall:narm}
%--------------------------------------------------

Table~\ref{tbl:overall_performance} shows that compared to the RNNs-based method \NARM, \method demonstrates superior performance on all the six datasets at all the evaluation metrics.
\NARM primarily uses RNNs to recurrently learn users' preferences.
As demonstrated in the literature~\cite{vaswani2017attention}, due to the recurrent nature, RNNs may struggle to 
model long-range dependencies within sequences, 
which could limit their ability 
to capture global information.
In \method, as illustrated in Figure~\ref{fig:star_graph}, 
the central internal node could  
aggregate information from every item node within one step.
Thus, compared to \NARM, \method could better capture long-range dependencies within sequences, 
and achieve substantial performance improvement on all the datasets. 
%

%--------------------------------------------------
\subsubsection{Comparison between \method and \HAM}
%unify comparing and comparison
\label{sec:results:overall:ham}
%--------------------------------------------------

Table~\ref{tbl:overall_performance} also presents that 
\method significantly outperforms the best-performing shallow method \HAM on all the six datasets.
Compared to \HAM, 
\method achieves a remarkable average improvement of 7.8\% and 6.0\% 
at Recall@$10$ and NDCG@$10$, respectively, across the six datasets.
\HAM employs the simple mean pooling to aggregate 
items and estimate users' intent.
Though efficient, this simple approach may not be 
sufficiently expressive to accurately estimate users' intent 
from their diverse interactions.
Different from \HAM, 
\method stacks multiple non-linear blocks to enable expressive models, which leads to a more accurate user intent modeling as compared to \HAM.

%--------------------------------------------------
\subsubsection{Comparison between \method and \SASRec}
\label{sec:results:overall:sasrec}
%--------------------------------------------------

Table~\ref{tbl:overall_performance} 
presents that both \method and \methodmu substantially outperform the best-performing SA-based method \SASRec.
Similarly to \method, \SASRec stacks multiple SA blocks to recursively 
capture users' intent, and has been demonstrated 
state-of-the-art performance~\cite{peng2021ham,sasrec} in SR.
However, as shown in Table~\ref{tbl:overall_performance}, \method consistently outperforms \SASRec on five out of six datasets (i.e., \Beauty, \Toys, \Children, \Comics and \MLOM).
Particularly, compared to \SASRec, in terms of Recall@$10$, \method achieves a significant average improvement of 14.2\% 
over the five datasets.
In terms of NDCG@$10$, \method also outperforms \SASRec with a significant average improvement of 15.1\% 
%at 95\% confidence interval 
over the five datasets.
Similarly, \methodmu also remarkably outperforms \SASRec across the six datasets with an average improvement of 7.0\% and 7.2\% at Recall@$10$ and NDCG@$10$, respectively.
%
%On \MLTM, although \method underperforms \SASRec, 
%its variant \methodmu could still achieve highly competitive performance with \SASRec.
%
From a graph perspective, \SASRec models sequences as fully connected graphs, which allow each item to aggregate information from other items within the sequence.
Though promising, this design results in quadratic time complexities and leads to the over-smoothing issue (Section~\ref{sec:results:smoothing}).
As shown in the literature~\cite{cai2020note,yang2020revisiting}, this issue could substantially limit the model scalability and degrade the recommendation performance.
In contrast, 
by modeling sequences as star graphs,
\method could fundamentally address the over-smoothing issue while still being able to capture the long-range dependencies within sequences.
Consequently, as presented in Table~\ref{tbl:overall_performance}, \method could enable significant improvement over \SASRec on different recommendation scenarios.
%

%--------------------------------------------------
\subsubsection{Comparison between \method and \SM}
\label{sec:results:overall:sm}
%--------------------------------------------------

%\bo{
Table~\ref{tbl:overall_performance} also shows that overall, both \method and \methodmu achieve superior performance over \SM on the six datasets.
For example, at Recall@10, 
\method and \methodmu achieves an average improvement of 11.4\% and 7.7\%, respectively, over the six datasets compared to \SM.
\SM mitigates over-smoothing by fusing embeddings from both shallow layers and deep layers 
using max pooling.
In contrast, \method fundamentally addresses 
over-smoothing by modeling sequences using star graphs, thereby avoiding information propagation among items.
The substantial improvement of \method over \SM shows that our approach is more effective than hierarchical fusion~\cite{shi2022revisiting} in both addressing over-smoothing, and improving recommendation performance.
%}

%--------------------------------------------------
\subsubsection{ Performance Summary across Datasets}
\label{sec:results:overall:summary}
%--------------------------------------------------

\input{tables/summary}
%\label{tbl:summary}

Table~\ref{tbl:summary} shows the average improvement of \methodmu and \method over the baseline methods \Caser, \HAM and \SASRec across the six datasets.
We focus on these baseline methods since they achieve the best performance in terms of at least one evaluation 
metric on the datasets.
As shown in Table~\ref{tbl:summary}, both \methodmu and \method significantly outperform the baseline methods across the six datasets.
For example, \methodmu and \method achieves a significant average improvement of 7.0\% and 10.6\%, respectively, over \SASRec at Recall@$10$ across the six datasets.
%
%Similarly, \method also outperforms \SASRec with a significant average improvement of 10.62\% at Recall@$10$ across the six datasets.

%**************************************************
\subsection{Comparison on 
Users with Different Activity Levels}
\label{sec:results:ufreq}
%**************************************************

\begin{figure}[!h]
	\vspace{-5pt}
	\centering
	\footnotesize
        \begin{footnotesize}
		\fbox{\begin{minipage}{.8\linewidth}
				\centering
				\blueline\quad  \method\quad\quad
                \magentaline\quad \methodmu\quad\quad
                \redline\quad \SASRec 
			\end{minipage}
		}
	\end{footnotesize}
        \\
	\begin{subfigure}{0.32\linewidth}
		\centering
		%\hspace*{20pt}
		\includegraphics[width=\linewidth]{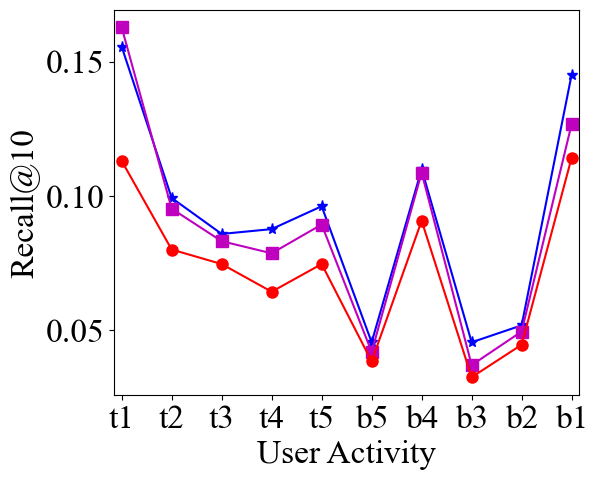}
		%\vspace*{10pt}
		\caption{\Beauty}
		\label{fig:ufreq:beauty}
	\end{subfigure}
	\begin{subfigure}{0.32\linewidth}
		\centering
		%\hspace*{20pt}
		\includegraphics[width=\linewidth]{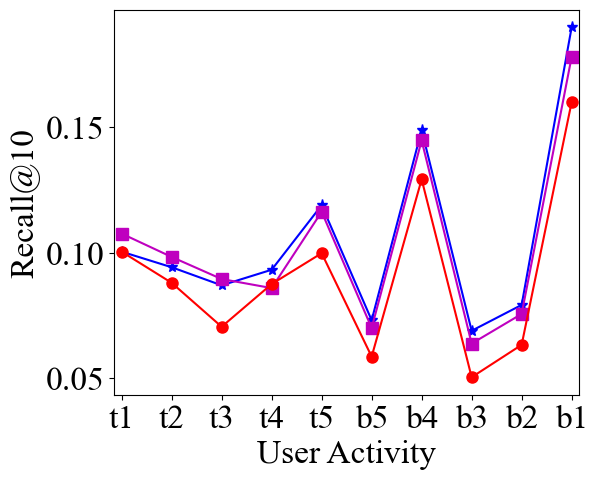}
		%\vspace*{10pt}
		\caption{\Toys}
		\label{fig:ufreq:toys}
	\end{subfigure}
	\begin{subfigure}{0.32\linewidth}
		\centering
		%\hspace*{20pt}
		\includegraphics[width=\linewidth]{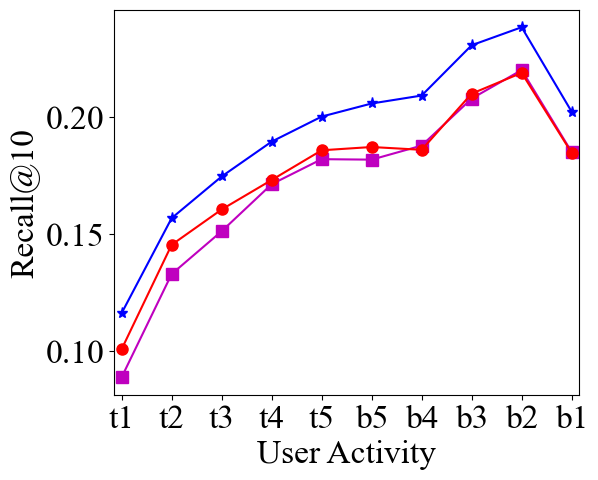}
		%\vspace*{10pt}
		\caption{\Children}
		\label{fig:ufreq:child}
	\end{subfigure}
	%
%	\begin{subfigure}{0.23\linewidth}
%		\centering
%		%\hspace*{20pt}
%		\includegraphics[width=.90\linewidth]{plots/activity/Comics_act.png}
%		%\vspace*{10pt}
%		\caption{\Comics}
%		\label{fig:ufreq:comics}
%	\end{subfigure}
%
\vspace{-10pt}
\caption{Performance on users of different activity levels}
\label{fig:ufreq}
\vspace{-10pt}
\end{figure}

We also compare the performance of \method and \methodmu with the best-performing SA-based SR method \SASRec 
on users with different activity levels.
Specifically, we quantify users' activity levels 
using the number of interactions in their history, 
and bin users into different buckets based on their activity levels.
In this analysis, we use ten buckets in total:
five buckets correspond to the top-10\%, %top 10-20\%, 
$\cdots$, top 40-50\% most active users; 
the other five buckets are for the bottom 40-50\%, 
%bottom 30-40\%, 
$\cdots$, bottom-10\% most active users.
%
%\bo{
Figure~\ref{fig:ufreq} presents the performance of \method, \methodmu and \SASRec on \Beauty, \Toys and \Children.
We present the results on \Comics, \MLOM and \MLTM in  Appendix~\ref{sec:more_results:ufreq} (Figure~\ref{fig:ufreq_more}).
The results on \Comics, \MLOM and \MLTM have a similar trend with that shown in Figure~\ref{fig:ufreq}.
%}
%
%We conduct this analysis using the widely used \Beauty, \Toys, \Children and \Comics datasets, 
%and present the performance of \method and \SASRec on users of different activity levels in Figure~\ref{fig:ufreq}.
%
%In this analysis, 
%We train \method, \methodmu and \SASRec using their best-performing hyper-parameters on these datasets.

As shown in Figure~\ref{fig:ufreq}, \method outperforms \SASRec by a remarkable margin on both active and less active users across the 
three datasets.
%
%\bo{
Similarly, \methodmu substantially outperforms \SASRec 
on users of different activity levels on \Beauty and \Toys, 
and achieves highly competitive performance with \SASRec on \Children.
%}
%
These results signify that by modeling sequences using star graphs, 
\method and \methodmu could more accurately estimate 
the intent of users with different activity levels 
compared to \SASRec.
Note that users' activity level 
is not the only factor 
in determining the difficulty of estimating their intent.
As a result, 
we do not expect a strictly 
increasing or decreasing performance curve 
as activity levels change.

%**************************************************
\subsection{Analysis on Over-smoothing}
\label{sec:results:smoothing}
%**************************************************

We analyze if the state-of-the-art SA-based SR method \SASRec suffers from the over-smoothing issue.
Specifically, for each dataset, 
we train a \SASRec model consisting of six SA blocks. 
Apart from the number of blocks, 
we use the best-performing hyper-parameters 
of \SASRec on each dataset %(\bo{Appendix}~\ref{sec:reproducibility}) 
to train the model.
Subsequently, 
%on each dataset, 
%\bo{
we calculate the average similarity between the embeddings of items within a sequence for each SA block.
We denote the average similarity calculated from the output of the $m$-th SA block as $a^{m}$.
We discuss the details on the calculation of $a^{m}$ in Appendix~\ref{sec:more_results:sim}.
%}
%

\begin{figure}
	%\vspace{-10pt}
	\centering
	\footnotesize
    \begin{footnotesize}
	\fbox{\begin{minipage}{.75\linewidth}
			\centering
			\blueline\quad  \method\quad\quad \redline\quad \SASRec 
			\end{minipage}
	}
	\end{footnotesize}
    \\
	\begin{subfigure}{0.32\linewidth}
		\centering
		\includegraphics[width=\linewidth]{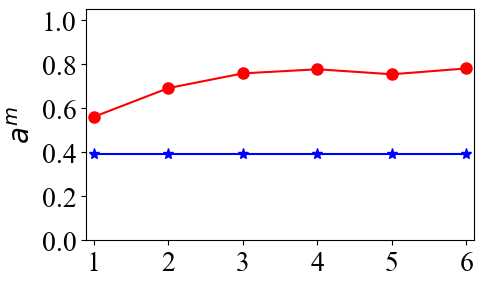}
		%\vspace*{10pt}
		\caption{\Beauty}
		\label{fig:over:beauty}
	\end{subfigure}
	\begin{subfigure}{0.32\linewidth}
		\centering
		%\hspace*{20pt}
		\includegraphics[width=\linewidth]{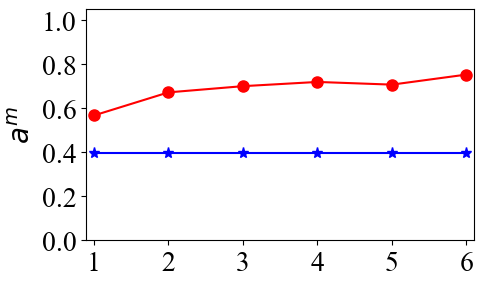}
		%\vspace*{10pt}
		\caption{\Toys}
		\label{fig:over:toys}
	\end{subfigure}
	\begin{subfigure}{0.32\linewidth}
		\centering
		%\hspace*{20pt}
		\includegraphics[width=\linewidth]{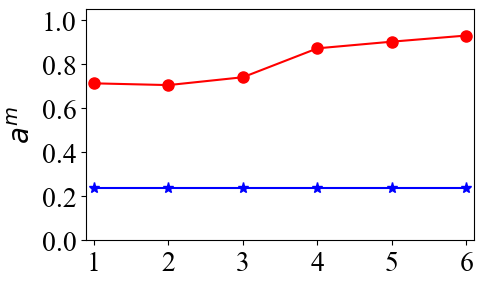}
		%\vspace*{10pt}
		\caption{\Children}
		\label{fig:over:child}
	\end{subfigure}
 \\
	\begin{subfigure}{0.32\linewidth}
		\centering
		%\hspace*{20pt}
		\includegraphics[width=\linewidth]{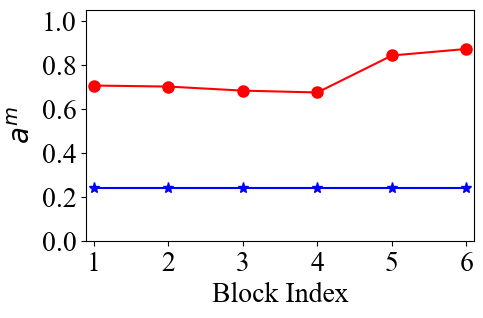}
		%\vspace*{10pt}
		\caption{\Comics}
		\label{fig:over:comics}
	\end{subfigure}
	\begin{subfigure}{0.32\linewidth}
		\centering
		%\hspace*{20pt}                    
		\includegraphics[width=\linewidth]{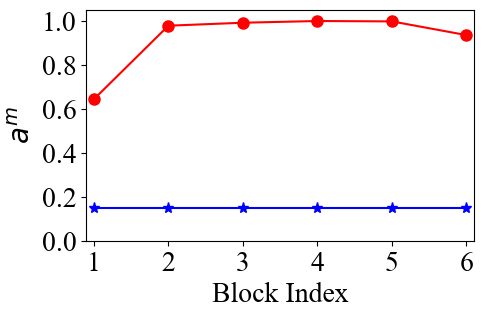}
		%\vspace*{10pt}                    
		\caption{\MLOM}
		\label{fig:over:ml1m}
	\end{subfigure}
	\begin{subfigure}{0.32\linewidth}
		\centering
		%\hspace*{20pt}                    
		\includegraphics[width=\linewidth]{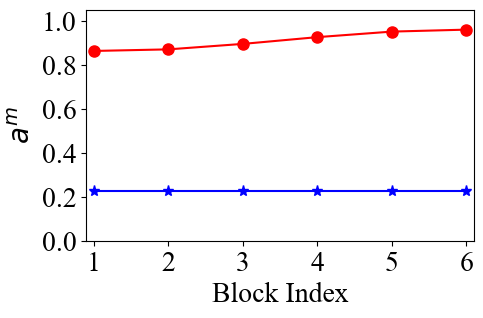}
		%\vspace*{10pt}                    
		\caption{\MLTM}
		\label{fig:over:ml20m}
	\end{subfigure}
\vspace{-10pt}
\caption{The average similarity $a^{m}$ in different blocks of \method and \SASRec}
\label{fig:over}
\vspace{-10pt}
\end{figure}

We use $a^{m}$ to assess whether the \SASRec model suffers from the over-smoothing issue.
%Subsequently, in the $m$-th block, 
%we calculate $a^{m}$ 
%as the average of $a_i^{m}$ over all the interaction sequences, 
%and use $a^{m}$ to assess whether the \SASRec model suffers from the over-smoothing issue.
%
In particular, 
the increase of $a^{m}$ over $m$ 
implies that generally, 
the embeddings of items
within a user's interaction sequence
become more and more similar across blocks. 
This serves as strong evidence 
to demonstrate that 
the \SASRec model suffers from 
the over-smoothing issue.
%
%In each dataset, we train a \method model using the same hyper-parameters as in the \SASRec model described above. 
%
%We then calculate $a^{m}$ for each block in order to assess whether the \method model suffers from the over-smoothing issue.
%
%To evaluate if \method is also affected 
%by the over-smoothing issue, 
%we train a \method model on each dataset 
%using the same hyper-parameters as in \SASRec.
%
%We then calculate $a^{m}$ from each block of the \method model and observe if $a^{m}$ increases over $m$.
%
We also evaluate if \method suffers from over-smoothing by calculating $a^{m}$ from the \method model trained using the same hyper-parameters as in \SASRec.
We present $a^{m}$ calculated from \method and \SASRec in Figure~\ref{fig:over}.
Note that, we observe a similar trend to that in Figure~\ref{fig:over} 
when using a smaller number of blocks.
Due to the space limit, 
we present results 
for using six blocks only.     

As illustrated in Figure~\ref{fig:over}, 
in \SASRec, 
$a^{m}$ increases across blocks 
on all the six datasets.
For example, on \Beauty, 
the average similarity rises 
from 0.56 (i.e., $a^1$) in the first block 
to 0.78 (i.e., $a^6$) in the last block.
These results demonstrate that \SASRec 
substantially suffers from the over-smoothing issue.
In contrast, 
as shown in Figure~\ref{fig:over}, 
$a^{m}$ from \method remains constant across blocks, 
indicating that \method is not affected by over-smoothing.
The key distinction between \method and \SASRec 
lies in their sequence modeling approaches.
While \SASRec utilizes fully connected graphs to model sequences, 
\method instead models sequences using star graphs.
As a result, 
\method does not require information propagation among item nodes, and 
allows each item to maintain its individual information across blocks.
Consequently, \method could fundamentally address the over-smoothing issue.
As shown in the literature~\cite{xu2018representation,shi2022revisiting}, 
the over-smoothing issue could substantially 
deteriorate the model scalability, 
and thus, degrade the model performance in the task of interest.
Therefore, by addressing this issue, 
\method could achieve superior performance over \SASRec on benchmark datasets as shown in Table~\ref{tbl:overall_performance}.
%

%**************************************************
\subsection{Analysis on Scalability}
\label{sec:results:scalability}
%**************************************************

The huge success of Transformer~\cite{vaswani2017attention} 
highlights the importance of a method's scalability in determining its utility for real-world applications. 
In this analysis, 
we evaluate the scalability of \method 
and \SASRec with respect to the number of blocks and the embedding dimensions.
Particularly, we assess the ability of \method and \SASRec in learning deep (i.e., with a large number of blocks) 
and wide (i.e., with a large embedding dimension) models on benchmark datasets.
To enable a fair comparison, 
in this analysis, 
we apply the best-performing 
hyper-parameters of \SASRec on each dataset 
for \SASRec and \method.
%

%--------------------------------------------------
%\subsubsection{Scalability over the number of blocks}
%\label{sec:results:scalability:layer}
%--------------------------------------------------

\begin{figure}[!h]
	\vspace{-10pt}
	\centering
	\footnotesize
        \begin{footnotesize}
		\fbox{\begin{minipage}{.75\linewidth}
				\centering
				\blueline\quad  \method\quad\quad \redline\quad \SASRec 
			\end{minipage}
		}
	\end{footnotesize}
    \\
	\begin{subfigure}{0.32\linewidth}
		\centering
		%\hspace*{20pt}
		\includegraphics[width=\linewidth]{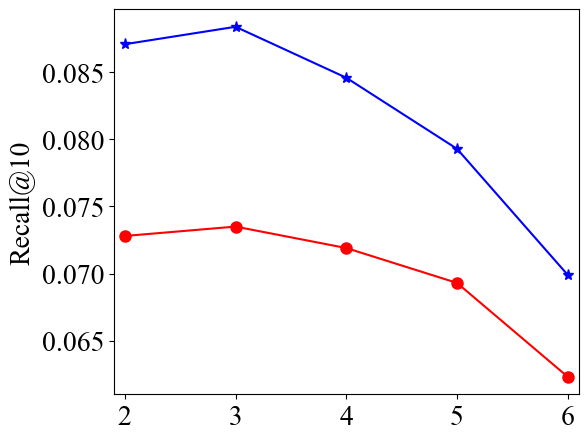}
		%\vspace*{15pt}
		\caption{\Beauty}
		\label{fig:sta:beauty}
	\end{subfigure}
	\begin{subfigure}{0.32\linewidth}
		\centering
		%\hspace*{20pt}
		\includegraphics[width=\linewidth]{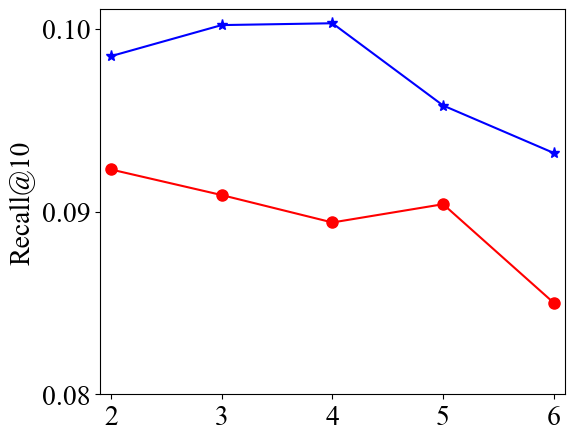}
		%\vspace*{15pt}
		\caption{\Toys}
		\label{fig:sta:toys}
	\end{subfigure}
	\begin{subfigure}{0.326\linewidth}
		\centering
		%\hspace*{20pt}                    
		\includegraphics[width=\linewidth]{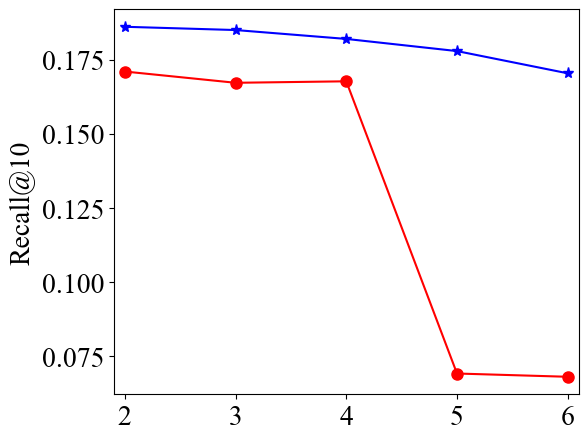}
		%\vspace*{15pt}                    
		\caption{\Children}
		\label{fig:sta:child}
	\end{subfigure}
    \\
	\begin{subfigure}{0.32\linewidth}
		\centering
		%\hspace*{20pt}
		\includegraphics[width=\linewidth]{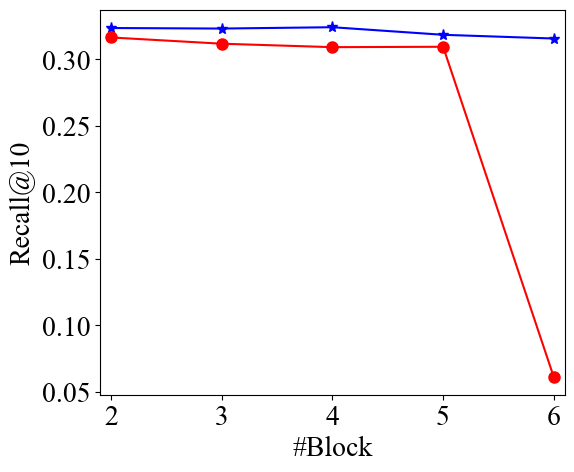}
		%\vspace*{15pt}
		\caption{\Comics}
		\label{fig:sta:comics}
	\end{subfigure}
	\begin{subfigure}{0.32\linewidth}
		\centering
		%\hspace*{20pt}                    
		\includegraphics[width=\linewidth]{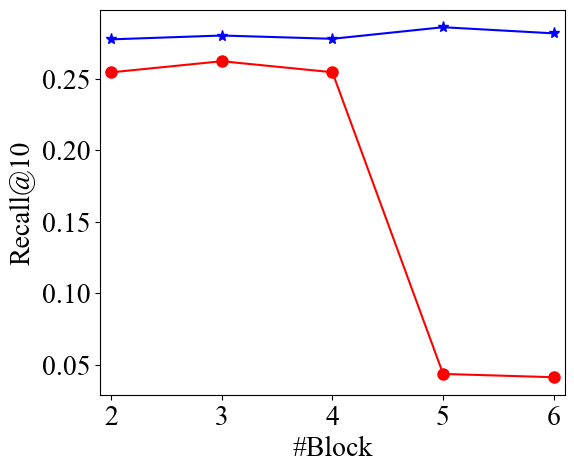}
		%\vspace*{15pt}
		\caption{\MLOM}
		\label{fig:sta:ml1m}
	\end{subfigure}
	\begin{subfigure}{0.326\linewidth}
		\centering
		%\hspace*{20pt}
		\includegraphics[width=\linewidth]{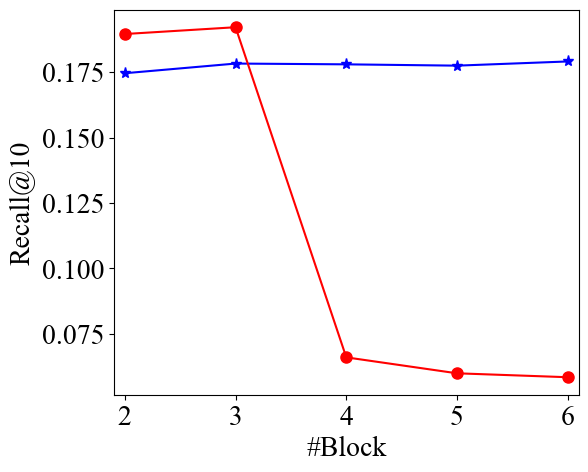}
		%\vspace*{15pt}
		\caption{\MLTM}
		\label{fig:sta:ml20m}
	\end{subfigure}
	%\end{minipage}
	%
	\vspace{-10pt}
	\caption{Performance over different numbers of blocks}
	\label{fig:sta}
	\vspace{-10pt}
\end{figure}

Figure~\ref{fig:sta} presents the performance of \SASRec and \method 
at Recall@$10$
over different numbers of blocks 
on the six benchmark datasets.
We limit the maximum number of blocks ($n_b$) to six, 
as we got the out-of-memory issue 
with \SASRec when $n_b>6$.

Figure~\ref{fig:sta} demonstrates that \method outperforms \SASRec at the scalability with respect to $n_b$. 
In particular, 
\method maintains reasonable performance as $n_b$ increases on all datasets, 
while \SASRec fails on four out of the six datasets 
(\Children, \Comics, \MLOM, and \MLTM) when $n_b$=$6$.
As discussed in Section~\ref{sec:results:smoothing}, 
the state-of-the-art SA-based SR method \SASRec substantially suffers from the over-smoothing issue.
Consequently, 
\SASRec struggles to learn deep models, 
as the item embeddings become increasingly similar across SA blocks~\cite{xu2018representation,zhao2019pairnorm,shi2022revisiting}.
In contrast, as shown in Section~\ref{sec:results:smoothing}, 
\method is not affected by over-smoothing, 
and thus, 
could enable superior scalability with respect to $n_b$ compared to \SASRec.
It is worth noting that, 
on \MLOM and \MLTM, 
the best performing \methodmu and \method models 
have at least 4 blocks (Appendix~\ref{sec:reproducibility}),
which reveals that 
the better scalability of \method at $n_b$ could 
translate into superior recommendation performance 
on real datasets.
%
%We notice that on \Toys, when $n_b$=$6$, \method considerably underperforms \SASRec. 
%
%However, it still achieves a reasonable Recall@$10$ of 0.0713.

%\bo{
Due to the space limit, we present the scalability comparison between \method and \SASRec in terms of embedding dimensions (i.e., $d$) in Appendix~\ref{sec:more_results:dim}.
As shown in Appendix~\ref{sec:more_results:dim} (Figure~\ref{fig:stad}), \method also substantially outperforms \SASRec at the scalability over $d$.
We refer the audience to Appendix~\ref{sec:more_results:dim} for a more detailed discussion.
\subsection{Comparison on Run-time Performance}
\label{sec:results:runtime}
%**************************************************

\begin{figure}[!h]
	\vspace{-5pt}
	\centering
	\footnotesize
        \begin{footnotesize}
		\fbox{\begin{minipage}{.75\linewidth}
				\centering
				\blueline\quad  \method\quad\quad \redline\quad \SASRec 
			\end{minipage}
		}
	\end{footnotesize}
        \\
	\begin{subfigure}{0.32\linewidth}
		\centering
		%\hspace*{20pt}
		\includegraphics[width=\linewidth]{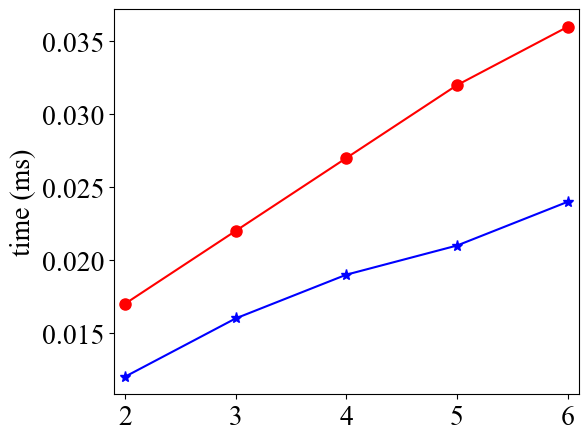}
		%\vspace*{10pt}
		\caption{\Beauty}
		\label{fig:runtime:beauty}
	\end{subfigure}
	\begin{subfigure}{0.32\linewidth}
		\centering
		%\hspace*{20pt}
		\includegraphics[width=\linewidth]{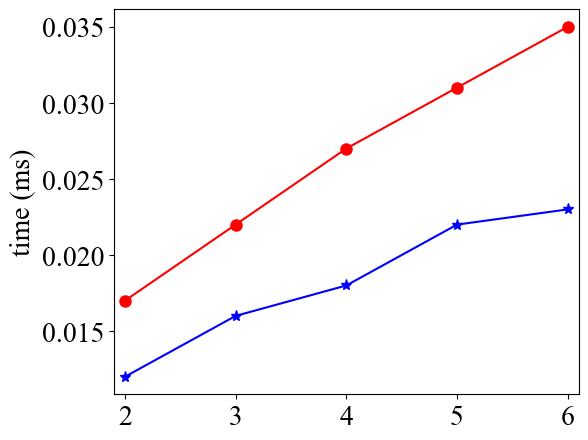}
		\caption{\Toys}
		\label{fig:runtime:toys}
	\end{subfigure}
	\begin{subfigure}{0.32\linewidth}
		\centering
		%\hspace*{20pt}
		\includegraphics[width=\linewidth]{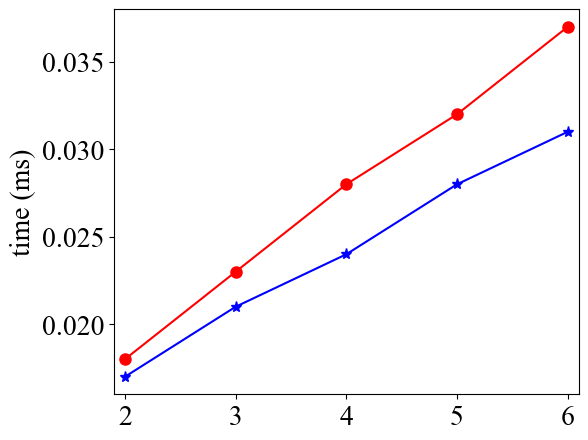}
		%\vspace*{10pt}
		\caption{\Children}
		\label{fig:runtime:child}
	\end{subfigure}
 \\
	\begin{subfigure}{0.32\linewidth}
		\centering
		%\hspace*{20pt}
		\includegraphics[width=\linewidth]{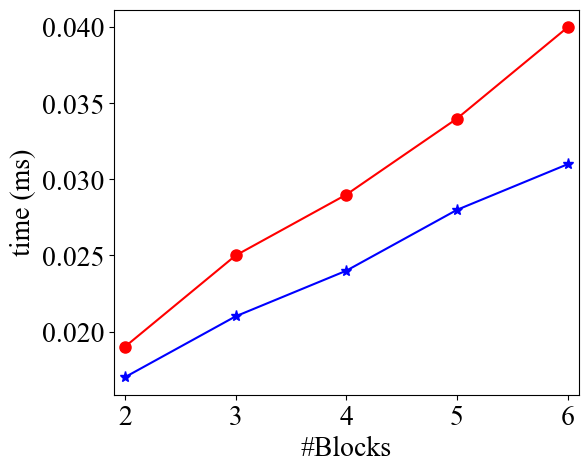}
		%\vspace*{10pt}
		\caption{\Comics}
		\label{fig:runtime:comics}
	\end{subfigure}
	\begin{subfigure}{0.32\linewidth}
		\centering
		%\hspace*{20pt}                    
		\includegraphics[width=\linewidth]{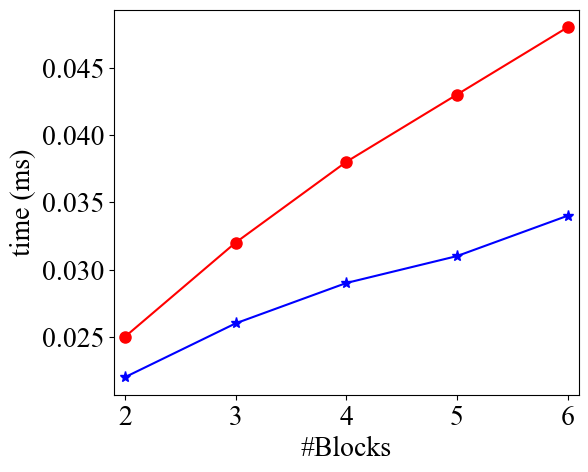}
		%\vspace*{10pt}                    
		\caption{\MLOM}
		\label{fig:runtime:ml1m}
	\end{subfigure}
	\begin{subfigure}{0.32\linewidth}
		\centering
		%\hspace*{20pt}                    
		\includegraphics[width=\linewidth]{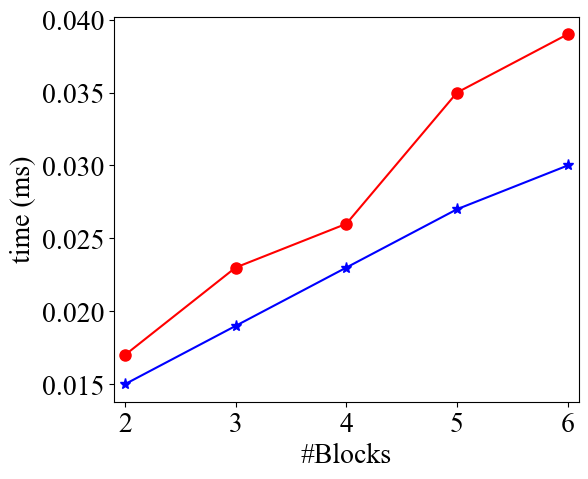}
		%\vspace*{10pt}                    
		\caption{\MLTM}
		\label{fig:runtime:ml20m}
	\end{subfigure}
\vspace{-10pt}
\caption{Run-time performance over the number of blocks}
\label{fig:runtime}
\vspace{-10pt}
\end{figure}

We conduct an analysis to evaluate the run-time performance of \method and \SASRec during testing. % on modern GPUs.
To enable a fair comparison, 
similarly to that in Section~\ref{sec:results:scalability}, 
we apply the best-performing hyper-parameters on \SASRec for \method and \SASRec, 
and compare their run-time performance during testing when using different numbers of blocks.
In addition, 
we perform the evaluation for both \method and \SASRec 
using NVIDIA Volta V100 GPUs, 
and report the 
average computation time per user over five runs 
in Figure~\ref{fig:runtime}.
Moreover, we use the same evaluation script 
for both \method and \SASRec to avoid 
any run-time performance differences 
that might arise from differences in the implementation.
We focus on the run-time performance during 
testing due to the fact that it
%the run-time performance in testing could 
could signify the models' latency in real-time recommendation, 
which could significantly affect the user experience and thus revenue.

As shown in Figure~\ref{fig:runtime}, 
%\xia{need to finish this section}
on all the datasets, 
\method achieves stronger run-time performance compared to \SASRec over different numbers of blocks, 
and the improvement increases as the number of blocks increases.
Particularly, 
when using the best-performing $n_b$ of \SASRec on each dataset
%when $n_b$ is the best-performing one for \SASRec 
%on each dataset 
(e.g., $n_b=2$ on \Beauty as in Appendix~\ref{sec:reproducibility}), 
%in terms of the run-time performance, 
\method achieves 
an average speedup of 16.6\% compared to \SASRec over the six datasets.
Note that, as shown in Figure~\ref{fig:sta}, 
with the same hyper-parameters, 
\method also outperforms \SASRec in terms of 
the recommendation performance on all the datasets except \MLTM.
These results demonstrate that compared to \SASRec, 
\method could generate more accurate recommendations in lower latency, 
thus significantly improving the user experience.
Note that on GPUs, all the computations are performed in parallel.
However, when calculating the time complexity (Section~\ref{sec:method:complexity}), 
we assume the computations are serial.
Therefore, in terms of the run-time 
performance on GPUs, 
the improvement of \method over \SASRec may 
not be as significant as that 
suggested by the time complexity.  
However, in many applications, the recommendation model could be 
deployed on edge devices with limited computing resources. 
In these applications, as suggested by the time complexity comparison (Section~\ref{sec:method:complexity}), 
\method could achieve a more substantial speedup 
over \SASRec.

%**************************************************
\subsection{Analysis on Attention Weights in \SASRec}
\label{sec:results:attention}
%**************************************************

\begin{figure}
	\vspace{-10pt}
	\includegraphics[width=.75\linewidth]{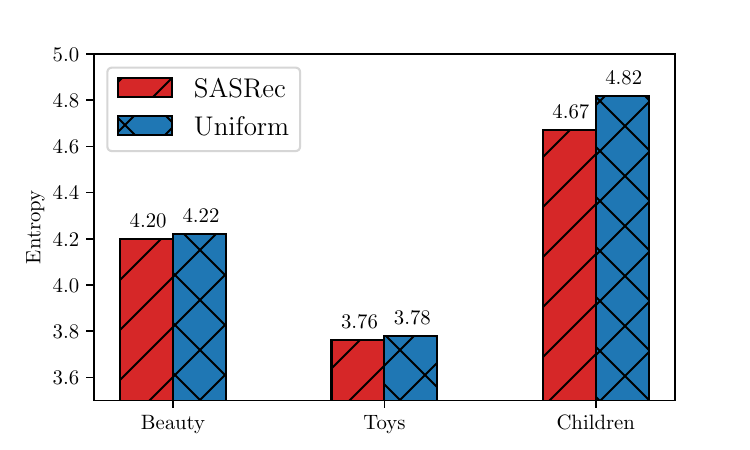}
    \vspace{-10pt}
	\caption{The average entropy of attention weight distribution in \SASRec and uniform distribution}
	\label{fig:entropy}
	\vspace{-10pt}
\end{figure}

%\bo{
We conduct an analysis to investigate the information gain derived from the attention weights learned in \SASRec.
Particularly, 
we measure the information gain by comparing the average entropy from the attention weight distributions and that from uniform distributions (i.e., weigh items equally).
We present more details on the calculation of the average entropy in Appendix~\ref{sec:more_results:entropy}.
A higher difference between the average entropies 
indicates a larger information gain derived from the learned attention weights.
Figure~\ref{fig:entropy} shows the average entropy from attention weight distributions and uniform distributions on \Beauty, \Toys and \Children.
Following \SASRec, we focus on the attention weights learned in the first SA block in this analysis.
%}

%\bo{
As shown in Figure~\ref{fig:entropy}, 
on all the three datasets, 
the difference between the average entropy 
from attention weight distributions 
and uniform distributions is highly marginal, 
indicating that \SASRec achieves limited information gain from the learned attention weights.
An important difference between \method and \SASRec is that 
\SASRec captures the dependencies in each pair of items by learning attention weights, while \method does not explicitly model these dependencies.
As shown in Figure~\ref{fig:entropy}, on the notoriously sparse recommendation data~\cite{peng2021ham}, limited information gain could be achieved by capturing the dependency in each item pair.
As a result, as shown in Table~\ref{tbl:overall_performance} and Table~\ref{tbl:summary}, without explicitly modeling dependencies in item pairs, \method and \methodmu could still achieve significant improvement over \SASRec on benchmark datasets.
%}

%%%%%%%%%%%%%%%%%%%%%%%%%%%%%%%%%%%%%%%%%%%%%%%%%%%
\section{Conclusion}
\label{sec:conclusion}
%%%%%%%%%%%%%%%%%%%%%%%%%%%%%%%%%%%%%%%%%%%%%%%%%%%

In this paper, 
we identify the over-smoothing issue 
in SA-based SR methods.
To address this issue, 
we model sequences using star graphs and develop \method for SR.
Different from SA-based methods in which each item could aggregate information from all others, 
\method introduces an internal node for information aggregation 
and does not propagate information among item nodes.
Consequently, \method could fundamentally address the over-smoothing issue and achieve
a linear time complexity 
with respect to the sequence length. 
%
%and enables each item to maintain its individual information across blocks to tackle the over-smoothing issue.
%
We extensively evaluate \method 
against 
ten state-of-the-art baseline methods on six benchmark datasets.
Our experimental results show that overall,  \method outperforms baseline methods on all the datasets except for \MLTM, with an improvement of up to 10.10\%.
On \MLTM, a variant of \method could still significantly outperform all the baseline methods.
Our analysis shows that \method 
achieves superior scalability 
with respect to both the number of blocks 
and the embedding dimensions 
compared to the state-of-the-art 
SA-based SR method \SASRec.
This improved scalability could lead to enhanced recommendation performance in benchmark datasets.
In addition, our complexity analysis and run-time performance comparison together demonstrate that 
\method is both theoretically and practically more efficient than \SASRec.
Thus, \method could be particularly 
desirable in applications with limited 
computing resources.
%
%\bo{
Our analysis also suggests that \SASRec achieves limited information gain by explicitly modeling dependencies between items using the SA mechanism.
Therefore, without modeling the dependencies, \method could still outperform \SASRec in benchmark datasets.
%}

%\begin{acks}
%To Robert, for the bagels and explaining CMYK and color spaces.
%\end{acks}


\clearpage

\bibliographystyle{ACM-Reference-Format}
\bibliography{main}

\clearpage

\appendix

\setcounter{table}{0}
\setcounter{figure}{0}
\renewcommand{\thetable}{A\arabic{table}}
\renewcommand{\thefigure}{A\arabic{figure}}

%%%%%%%%%%%%%%%%%%%%%%%%%%%%%%%%%%%%%%%%%%%%%%%%%%%
\section{Notations}
\label{sec:not}
%%%%%%%%%%%%%%%%%%%%%%%%%%%%%%%%%%%%%%%%%%%%%%%%%%%

\input{tables/notations}

%\label{tbl:notations}

Table~\ref{tbl:notations} summarizes the key notations used in this paper.

%%%%%%%%%%%%%%%%%%%%%%%%%%%%%%%%%%%%%%%%%%%%%%%%%%%
\section{Reproducibility}
\label{sec:reproducibility}
%%%%%%%%%%%%%%%%%%%%%%%%%%%%%%%%%%%%%%%%%%%%%%%%%%%

\input{tables/parameters}

We implement \method and \methodmu in Python 3.9.13 with PyTorch 1.10.2.
%~\footnote{\url{https://pytorch.org}}.
We use Adam optimizer with learning rate 1e-3 
for \method and \methodmu on all the datasets.
%
%We decay the learning rate by 0.995 after each epoch.
%
For all the baseline methods except \FPMC and \BERT, 
we use the implementation provided by the authors in
GitHub.
For \FPMC and \BERT, we use the implementation in RecBole~\footnote{\url{https://recbole.io/}}, 
a widely used library to benchmark recommendation methods.
For \SASRec, \FMLP, \methodmu and \method, 
we search the embedding dimension $d$ in $\{64, 128, 256, 512\}$, 
the length of the fixed-length sequences $n$ in $\{50, 75, 100, 125, 150, 175, 200\}$, 
the number of heads $n_h$ in $\{1, 2, 4, 8, 16\}$, 
and the number of blocks $n_b$ in $\{1, 2, 3, 4, 5\}$.
For \BERT, we search $d$ in $\{64, 128, 256, 512\}$, 
$n_h$ in $\{1, 2, 4, 8, 16\}$ 
and $n_b$ in $\{1, 2, 3, 4, 5\}$.
%
%\bo{
For \SC and \SM, 
the search range of $d$, $n$ and $n_h$ is the same with that in \SASRec.
We search $n_b$ in \SC and \SM in $\{2, 3, 4, 5\}$ as 
\SC and \SM are equivalent to \SASRec when $n_b$=$1$.
%}
%
We use GELU~\cite{hendrycks2016gaussian} as 
the activation function in the feed-forward layer for 
\SASRec, \SC, \SM \methodmu and \method.
We use the PyTorch implementation in GitHub~\footnote{\url{https://github.com/pmixer/SASRec.pytorch}}
for \SASRec.
For \FPMC, we search $d$ in $\{64, 128, 256, 512\}$.
For \Caser
%~\footnote{\url{https://github.com/graytowne/caser_pytorch}}, 
we search $d$ in $\{64, 128, 256, 512\}$,
the length of the subsequences $n_s$ in $\{4, 5, 6\}$,  
the number of negative items during training $n_p$ in $\{1, 2\}$, 
the number of vertical filters in CNNs $n_v$ in $\{1, 2, 4\}$, 
and the number of horizontal filters in CNNs $n_f$ in $\{4, 8, 16\}$.
For \NARM
%~\cite{narm_pytorch},
we search $d$ in $\{64 ,128, 256, 512\}$ and 
the learning rate $l_r$ in \{1e-2, 1e-3, 1e-4\}. 
For \HGN
%~\cite{hgn_pytorch},
%~\footnote{\mbox{\url{https://github.com/allenjack/HGN}}}, 
we search $d$ in $\{64, 128, 256, 512\}$, 
$n_s$ in $\{3, 4, 5\}$,  
$n_p$ in $\{1, 2\}$, 
and the regularization factor $\lambda$ in \{0, 1e-3, 1e-4\}.
For \HAM
%~\cite{ham_pytorch},
we search $d$ in $\{64, 128, 256, 512\}$, 
$n_s$ in $\{3, 4, 5\}$,  
$n_p$ in $\{1, 2\}$, 
$\lambda$ in \{0, 1e-3, 1e-4\}, 
the number of items in low order $n_l$ in $\{1, 2, 3\}$,
and the order of item synergies $n_o$ in $\{1, 2, 3\}$. 
%
%For each baseline method except \FPMC and \BERT, we use the training objective in their original implementations.
%
Following the instruction in RecBole, 
we use the Bayesian personalized ranking loss~\cite{bpr} 
and cross-entropy loss for \FPMC and \BERT, respectively.
We report the best-performing hyper-parameters 
of \method, \methodmu and baseline methods in Table~\ref{tbl:para}.

%%%%%%%%%%%%%%%%%%%%%%%%%%%%%%%%%%%%%%%%%%%%%%%%%%%
\section{Dataset Pre-processing and Statistics}
\label{sec:datasets_statistics}
%%%%%%%%%%%%%%%%%%%%%%%%%%%%%%%%%%%%%%%%%%%%%%%%%%%

\input{tables/dataset.tex}

Following \SASRec, for \Beauty, \Toys and \MLOM, 
we only keep the users and items with at least 5 ratings.
For \Children, \Comics and \MLTM, which are not used in \SASRec, 
following \HAM, 
we keep users with at least 10 ratings, and items with at least 5 ratings. 
%\xia{instead of "remove", use "keep"}
%
Following the literature~\cite{sasrec,peng2021ham}, 
we consider the ratings as users' implicit feedback, 
and convert ratings into binary values.
Particularly, for ratings with a range from 1 to 5, 
we convert ratings 4 and 5 to binary value 1, or 0 otherwise.
Table~\ref{tbl:dataset} presents the statistics of the 
processed datasets.

%%%%%%%%%%%%%%%%%%%%%%%%%%%%%%%%%%%%%%%%%%%%%%%%%%%
\section{More Experimental Results and Details}
\label{sec:more_results}
%%%%%%%%%%%%%%%%%%%%%%%%%%%%%%%%%%%%%%%%%%%%%%%%%%%

%**************************************************
\subsection{Comparison on 
Users with Different Activity Levels (Cont.)}
\label{sec:more_results:ufreq}
%**************************************************

\begin{figure}
	%\vspace{-10pt}
	\centering
	\footnotesize
        \begin{footnotesize}
		\fbox{\begin{minipage}{.8\linewidth}
				\centering
				\blueline\quad  \method\quad\quad
                \magentaline\quad \methodmu\quad\quad
                \redline\quad \SASRec 
			\end{minipage}
		}
	\end{footnotesize}
        \\
	\begin{subfigure}{0.32\linewidth}
		\centering
		%\hspace*{20pt}
		\includegraphics[width=\linewidth]{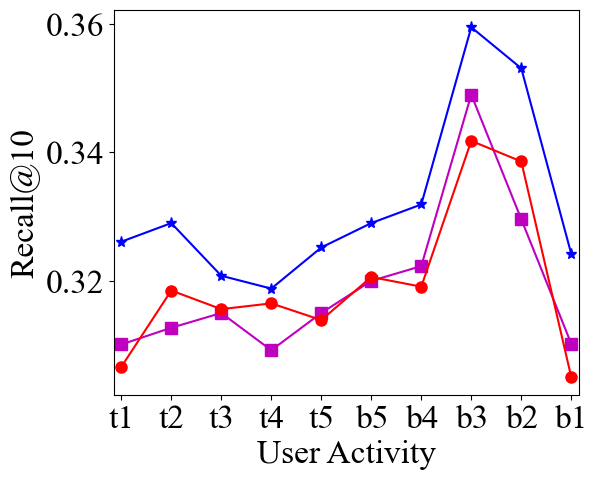}
		%\vspace*{10pt}
		\caption{\Comics}
		\label{fig:ufreq:comics}
	\end{subfigure}
	\begin{subfigure}{0.32\linewidth}
		\centering
		%\hspace*{20pt}
		\includegraphics[width=\linewidth]{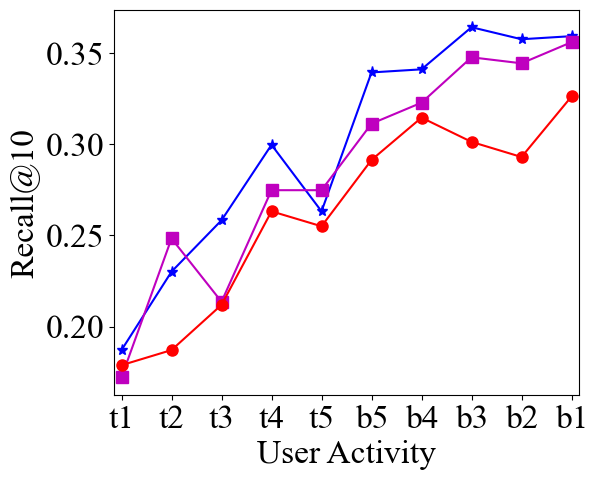}
		%\vspace*{10pt}
		\caption{\MLOM}
		\label{fig:ufreq:ml1m}
	\end{subfigure}
	\begin{subfigure}{0.32\linewidth}
		\centering
		%\hspace*{20pt}
		\includegraphics[width=\linewidth]{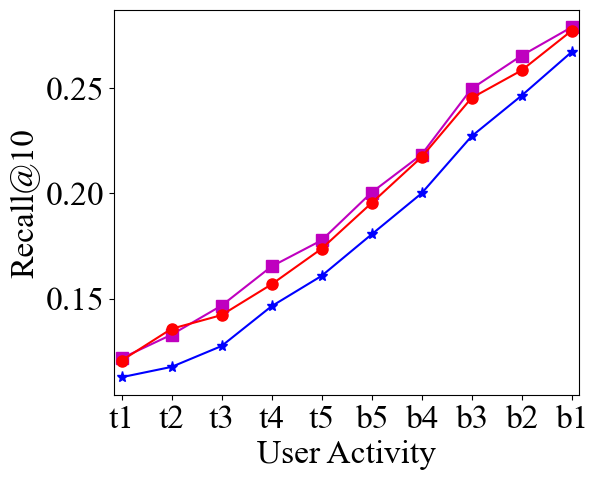}
		%\vspace*{10pt}
		\caption{\MLTM}
		\label{fig:ufreq:ml20m}
	\end{subfigure}
	%
%	\begin{subfigure}{0.23\linewidth}
%		\centering
%		%\hspace*{20pt}
%		\includegraphics[width=.90\linewidth]{plots/activity/Comics_act.png}
%		%\vspace*{10pt}
%		\caption{\Comics}
%		\label{fig:ufreq:comics}
%	\end{subfigure}
%
%\vspace{-10pt}
\caption{Performance on users of different activity levels}
\label{fig:ufreq_more}
%\vspace{-10pt}
\end{figure}

%\bo{
Figure~\ref{fig:ufreq_more} shows the performance of \method, \methodmu and \SASRec in users of different activity levels on the \Comics, \MLOM and \MLTM datasets.
As shown in Figure~\ref{fig:ufreq_more}, the performance on \Comics, \MLOM and \MLTM has a similar trend with that on \Beauty, \Toys and \Children (Figure~\ref{fig:ufreq} in Section~\ref{sec:results:ufreq}).
Particularly, \method substantially outperforms \SASRec on \Comics and \MLOM in both active and less active users.
Similarly, in most activity levels, \methodmu considerably outperforms \SASRec on \MLOM and \MLTM.
On \Comics, \methodmu also achieves highly comparable performance with \SASRec.
%}

%**************************************************
\subsection{Calculating Average Similarities Between Item embeddings}
\label{sec:more_results:sim}
%**************************************************

Given the transformed sequence of user $u_i$:
$B_i$=$\{v_{b_1}(i), v_{b_2}(i), \dots\}$ (Section~\ref{sec:method:seq}),  
we calculate the average similarity of the embeddings of items within a sequence for each SA block as follows:
\begin{equation}
    \label{eqn:avg}
    a_i^{m} = 
    \frac{1}{|B_i|^2} \sum_{j=1}^{|B_i|} \sum_{k=1}^{|B_i|} 
    \cos(\mathbf{e}_{j}^{m}, \mathbf{e}_{k}^{m}),  
\end{equation}
where $|B_i|$ is the length of $B_i$; 
$\mathbf{e}_{j}^{m}$ is the embedding of the $j$-th item in $B_i$ after 
the $m$-th SA block; 
$\cos(\cdot)$ is the cosine similarity;
$a_i^{m}$ is the average similarity between the embeddings of items in $B_i$ after the $m$-th SA block.
Note that as illustrated in Figure~\ref{fig:clique_graph}, 
\SASRec updates the embedding of each item within the sequence in each SA block.
Thus, we expect different embeddings 
for the same item in different blocks.
We eliminate $i$ in $\mathbf{e}_{j}^{m}(i)$ in Equation~\ref{eqn:avg} for simplicity.

Given $a_i^{m}$, we average it over all 
the users in the dataset:
\begin{equation}
    \label{avgavg}
    a^{m} = \frac{1}{|\mathbb{U}|} 
    \sum_{u_i \in \mathbb{U}} a_i^{m},  
\end{equation}
where $a^{m}$ is the average similarity over all the users and $\mathbb{U}$ is the set of all the users.

%**************************************************
\subsection{Scalability over Embedding Dimensions}
\label{sec:more_results:dim}
%**************************************************

\begin{figure}
	\vspace{0pt}
	\centering
	\footnotesize
         \begin{footnotesize}
		\fbox{\begin{minipage}{.75\linewidth}
				\centering
				\blueline\quad  \method\quad\quad \redline\quad \SASRec 
			\end{minipage}
		}
	\end{footnotesize}
        \\
	\begin{subfigure}{0.32\linewidth}
		\centering
		%\hspace*{20pt}                    
		\includegraphics[width=\linewidth]{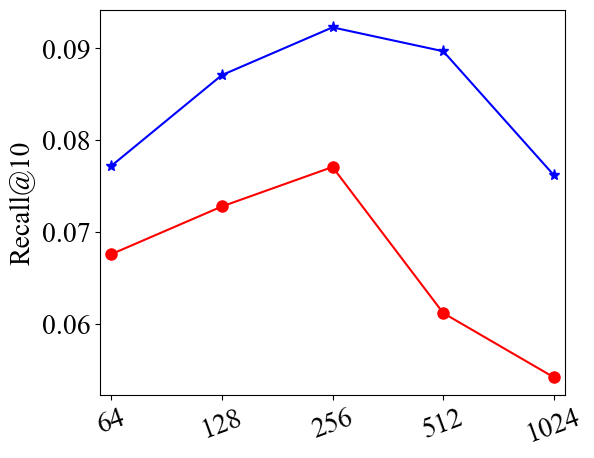}
		%\vspace*{20pt}                    
		\caption{\Beauty}
		\label{fig:stad:beauty}
	\end{subfigure}
	\begin{subfigure}{0.32\linewidth}
		\centering
		%\hspace*{20pt}                    
		\includegraphics[width=\linewidth]{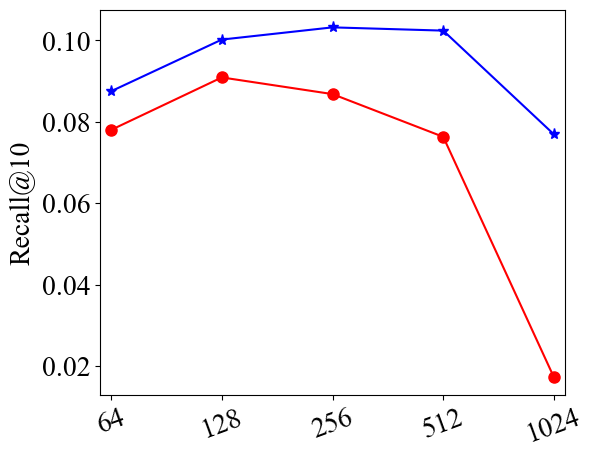}
		%\vspace*{20pt}                    
		\caption{\Toys}
		\label{fig:stad:toys}
	\end{subfigure}
	\begin{subfigure}{0.32\linewidth}
		\centering
		%\hspace*{20pt}                    
		\includegraphics[width=\linewidth]{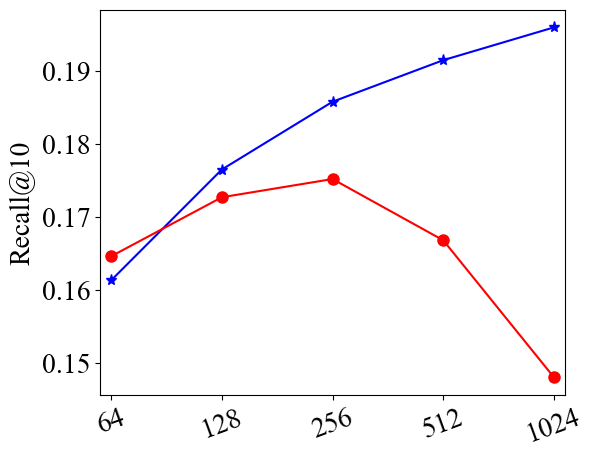}
		%\vspace*{20pt}                    
		\caption{\Children}
		\label{fig:stad:child}
	\end{subfigure}
	\\
	\begin{subfigure}{0.32\linewidth}
		\centering
		%\hspace*{20pt}
		\includegraphics[width=\linewidth]{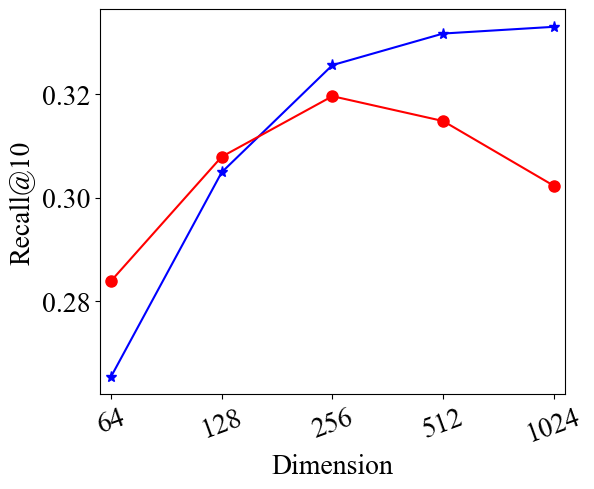}
		%\vspace*{20pt}
		\caption{\Comics}
		\label{fig:stad:comics}
	\end{subfigure}
	\begin{subfigure}{0.32\linewidth}
		\centering
		%\hspace*{20pt}                    
		\includegraphics[width=\linewidth]{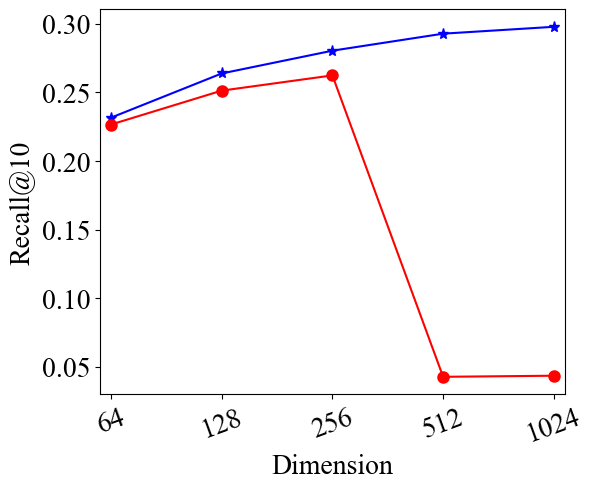}
		%\vspace*{20pt}
		\caption{\MLOM}
		\label{fig:stad:ml1m}
	\end{subfigure}
	\begin{subfigure}{0.32\linewidth}
		\centering
		%\hspace*{20pt}
		\includegraphics[width=\linewidth]{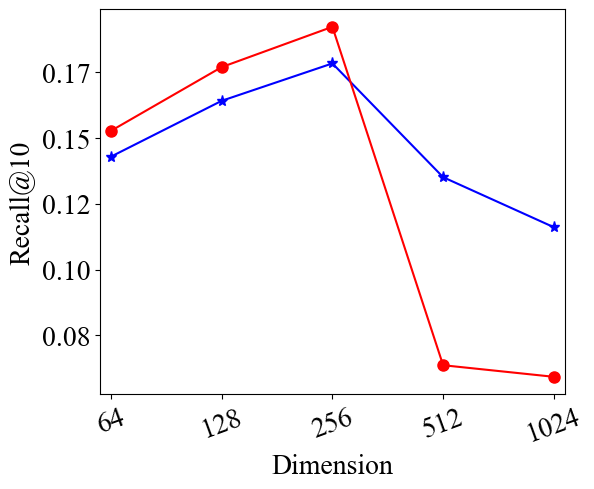}
		%\vspace*{20pt}
		\caption{\MLTM}
		\label{fig:stad:ml20m}
	\end{subfigure}
	\vspace{-10pt}
	\caption{Performance over different embedding dimensions}
	\label{fig:stad}
	\vspace{-10pt}
\end{figure}

Figure~\ref{fig:stad} presents the performance 
of \SASRec and \method at Recall@$10$
over different embedding dimensions 
on the six datasets.
We observe that when the embedding dimension $d$=$2048$, 
both \SASRec and \method were unable to complete training within 24 hours.
Thus, we limit the maximum embedding dimension to $1024$ in this analysis.
As shown in Figure~\ref{fig:stad}, 
\method is also more scalable than \SASRec with respect to $d$.
On all the datasets, 
\method could perform reasonably well when $d$=$1024$.
However, 
\SASRec fails on three out of the six datasets 
(\Toys, \MLOM and \MLTM) when $d$=$1024$.
These results demonstrate the superior scalability of \method over \SASRec with respect to $d$.
We also notice that as illustrated in Figure~\ref{fig:stad}, 
the recommendation performance of \method improves as $d$ increases on both \Children and \Comics.
These results suggest that better scalability in $d$ also leads to improved recommendation performance in real-world datasets.

%**************************************************
\subsection{Calculating Entropy 
from Attention Weight Distributions}
\label{sec:more_results:entropy}
%**************************************************

%\bo{
Given the attention map $A$ from the 
first SA block in \SASRec,  
we calculate the average entropy of 
the attention weight distributions
over all the items
as follows:
\begin{equation}
    \label{eqn:entropy}
    \frac{-1}{\sum \limits_{u_i \in \mathbb{U}} \sum \limits_{j=1}^n \mathbbm{1} (v_{b_j}(i) \ne v_0)} \sum \limits_{u_i \in \mathbb{U}} \sum \limits_{j=1}^{n} \mathbbm{1} (v_{b_j}(i) \ne v_0) \sum \limits_{k=1}^{j} (A_{ijk} \log(A_{ijk})), 
\end{equation}
where $\mathbbm{1}(x)$ is an indicator function (i.e., $\mathbbm{1}(x) = 1$ if $x$ is
true, otherwise 0);
$n$ is the length of the transformed sequence (Section~\ref{sec:method:seq});
$v_{b_j}(i) \ne v_0$ excludes the padding items;
and $-\sum \limits_{k=1}^{j} (A_{ijk} \log(A_{ijk}))$ is the entropy calculated from the attention weights of $v_{b_j}(i)$.
%}

\end{document}

%% file: define.tex
\newcommand{\method}{\mbox{$\mathop{\mathtt{MSSG}}\limits$}\xspace}
\newcommand{\methodmu}{\mbox{$\mathop{\mathtt{MSSG\text{-}u}}\limits$}\xspace}
\newcommand{\FPMC}{\mbox{$\mathop{\mathtt{FPMC}}\limits$}\xspace}
\newcommand{\BERT}{\mbox{$\mathop{\mathtt{Bert4Rec}}\limits$}\xspace}
\newcommand{\SASRec}{\mbox{$\mathop{\mathtt{SASRec}}\limits$}\xspace}
\newcommand{\Caser}{\mbox{$\mathop{\mathtt{Caser}}\limits$}\xspace}
\newcommand{\HGN}{\mbox{$\mathop{\mathtt{HGN}}\limits$}\xspace}
\newcommand{\HAM}{\mbox{$\mathop{\mathtt{HAM}}\limits$}\xspace}
\newcommand{\NARM}{\mbox{$\mathop{\mathtt{NARM}}\limits$}\xspace}
\newcommand{\GRURec}{\mbox{$\mathop{\mathtt{GRU4REC}}\limits$}\xspace}

\newcommand{\NextItRec}{\mbox{$\mathop{\mathtt{NextItRec}}\limits$}\xspace}
\newcommand{\FMLP}{\mbox{$\mathop{\mathtt{FMLP}}\limits$}\xspace}
\newcommand{\SGNN}{\mbox{$\mathop{\mathtt{SGNN\text{-}HN}}\limits$}\xspace}
\newcommand{\Concat}{\mbox{$\mathop{\mathtt{Concat Fusion}}\limits$}\xspace}
\newcommand{\Max}{\mbox{$\mathop{\mathtt{Max Fusion}}\limits$}\xspace}

\newcommand{\SC}{\mbox{$\mathop{\mathtt{SASRec\text{-}Cat}}\limits$}\xspace}
\newcommand{\SM}{\mbox{$\mathop{\mathtt{SASRec\text{-}Max}}\limits$}\xspace}

\newcommand{\Beauty}{\mbox{$\mathop{\texttt{Beauty}}\limits$}\xspace}
\newcommand{\Toys}{\mbox{$\mathop{\texttt{Toys}}\limits$}\xspace}
\newcommand{\Children}{\mbox{$\mathop{\texttt{Children}}\limits$}\xspace}
\newcommand{\Comics}{\mbox{$\mathop{\texttt{Comics}}\limits$}\xspace}
\newcommand{\MLOM}{\mbox{$\mathop{\texttt{ML-1M}}\limits$}\xspace}
\newcommand{\MLTM}{\mbox{$\mathop{\texttt{ML-20M}}\limits$}\xspace}

%% file: tables/overall.tex
\begin{table*}[!t]
\footnotesize
  \caption{Overall Performance}
  \centering
  \vspace{-10pt}
  \label{tbl:overall_performance}
  \begin{threeparttable}
      \begin{tabular}{
        @{\hspace{0pt}}l@{\hspace{5pt}}
	  @{\hspace{5pt}}l@{\hspace{5pt}}
		@{\hspace{6pt}}c@{\hspace{6pt}}
		@{\hspace{6pt}}c@{\hspace{6pt}}
		@{\hspace{6pt}}c@{\hspace{6pt}}
	@{\hspace{6pt}}c@{\hspace{6pt}}
        @{\hspace{6pt}}c@{\hspace{6pt}}
        @{\hspace{6pt}}c@{\hspace{4pt}}
        @{\hspace{2pt}}c@{\hspace{2pt}}
        @{\hspace{2pt}}c@{\hspace{2pt}}
        @{\hspace{2pt}}c@{\hspace{6pt}}
        @{\hspace{6pt}}c@{\hspace{6pt}}
        @{\hspace{6pt}}c@{\hspace{6pt}}
        @{\hspace{6pt}}c@{\hspace{2pt}}
        @{\hspace{2pt}}r@{\hspace{0pt}}
      }
      \toprule
      %move \NARM to \NARM & \HAM & \HGN
      Dataset & Metric & \FPMC & \Caser & \NARM & \HGN & \HAM & \SASRec & \SC & \SM 
      & \BERT & \FMLP & \methodmu 
      & \method & imprv (\%)\\
      \midrule
      \multirow{4}{*}{\Beauty}
      & Recall@10 & 0.0750 & 0.0647 & 0.0554 & 0.0782 & \underline{0.0865} & 0.0728 
      & 0.0704 & 0.0741 
      & 0.0549 & 0.0622 & 0.0874 & \textbf{0.0923} 
      & 6.71$^*$\\
      %0.0816 
      %& \textbf{0.0867} & 0.23\textcolor{white}{$^*$}\\
      & Recall@20 & 0.1086 & 0.0869 & 0.0835 & 0.1121 & \underline{0.1221} & 0.1083 
      & 0.1024 & 0.1106
      & 0.0776 & 0.0952 & 0.1249 & \textbf{0.1286} & 5.32$^*$\\
      %0.1168 & \textbf{0.1250} & 2.38\textcolor{white}{$^*$}\\
      \cline{2-15}
      & NDCG@10 & 0.0430 & 0.0371 & 0.0291 & 0.0434 
      & \underline{0.0493} & 0.0398
      & 0.0390 & 0.0397
      & 0.0298 & 0.0330 & 0.0486
      & \textbf{0.0516} & 4.67$^*$\\
      %& 0.0456 & \textbf{0.0493} & 0.00\textcolor{white}{$^*$}\\
      & NDCG@20 & 0.0514 & 0.0427 & 0.0362 & 0.0520 
      & \underline{0.0582} & 0.0487 
      & 0.0470 & 0.0489
      & 0.0355 & 0.0413 & 0.0581 
      & \textbf{0.0607} & 4.30$^*$\\
      %0.0545 & \textbf{0.0590} & 1.37\textcolor{white}{$^*$}\\
      \midrule
      \multirow{4}{*}{\Toys}
      & Recall@10 & 0.0869 & 0.0675 & 0.0557 & 0.0926 
      & \underline{0.1010} & 0.0909
      & 0.0843 & 0.0903
      & 0.0539 & 0.0787 & 0.1031 & \textbf{0.1055} & 4.46$^*$\\
      %0.0938 & \textbf{0.1016} & 0.59\textcolor{white}{$^*$}\\
      & Recall@20 & 0.1173 & 0.0895 & 0.0816 & 0.1240 
      & \underline{0.1303} & 0.1278 
      & 0.1123 & 0.1240
      & 0.0756 & 0.1072 & 0.1392 & \textbf{0.1434} & 10.10$^*$\\
      %0.1291 & \textbf{0.1374} & 5.45$^*$\\
      \cline{2-15}
      & NDCG@10 & 0.0517 & 0.0396 & 0.0308 & 0.0540 
      & \underline{0.0620} & 0.0521 
      & 0.0496 & 0.0515
      & 0.0290 & 0.0444 & 0.0604 & \textbf{0.0625} & 0.81\textcolor{white}{$^*$}\\
      %0.0539 & \textbf{0.0588} & -5.16$^*$\\
      & NDCG@20 & 0.0594 & 0.0452 & 0.0373 & 0.0619 
      & \underline{0.0694} & 0.0614 
      & 0.0566 & 0.0600
      & 0.0344 & 0.0516 & 0.0696 & \textbf{0.0720} & 3.75$^*$\\
      %0.0628 & \textbf{0.0679} & -2.16\textcolor{white}{$^*$}\\
      \midrule
      \multirow{4}{*}{\Children}
      & Recall@10 & 0.1298 & 0.1303 & 0.1126 & 0.1536 
      & 0.1729 & \underline{0.1752} 
      & 0.1471 & 0.1706
      & 0.1435 & 0.1675 & 0.1709 & \textbf{0.1924} & 9.82$^*$\\
      %& 0.1729 & \textbf{0.1901} & 8.50$^*$\\
      & Recall@20 & 0.1843 & 0.1832 & 0.1735 & 0.2118 
      & \underline{0.2366} & 0.2337 
      & 0.2064 & 0.2320
      & 0.1930 & 0.2249 & 0.2334 & \textbf{0.2589} & 9.43$^*$\\
      %& 0.2352 & \textbf{0.2578} & 8.96$^*$\\
      \cline{2-15}
      & NDCG@10 & 0.0763 & 0.0764 & 0.0585 & 0.0917 
      & 0.1038 & \underline{0.1086} 
      & 0.0861 & 0.1037
      & 0.0884 & 0.1040 & 0.1038 & \textbf{0.1176} & 8.29$^*$\\
      %& 0.1050 & \textbf{0.1162} & 7.00$^*$\\
      & NDCG@20 & 0.0900 & 0.0897 & 0.0738 & 0.1063 
      & 0.1199 & \underline{0.1234} 
      & 0.1010 & 0.1191
      & 0.1009 & 0.1184 & 0.1195 & \textbf{0.1343} & 8.83$^*$\\
      %& 0.1206 & \textbf{0.1332} & 7.94$^*$\\
      \midrule
      \multirow{4}{*}{\Comics}
      & Recall@10 & 0.2517 & 0.2320 & 0.1304 & 0.2857 
      & 0.3055 & \underline{0.3196} 
      & 0.2975 & 0.3144
      & 0.2363 & 0.3163 & 0.3193 & \textbf{0.3317} & 3.79$^*$\\
      %& 0.3169 & \textbf{0.3297} & 3.16$^*$\\
      & Recall@20 & 0.2986 & 0.2791 & 0.1896 & 0.3322 
      & 0.3513 & \underline{0.3650} 
      & 0.3456 & 0.3649
      & 0.2846 & 0.3631 & 0.3680 & \textbf{0.3833} & 5.01$^*$\\
      %& 0.3657 & \textbf{0.3823} & 4.74$^*$\\
      \cline{2-15}
      & NDCG@10 & 0.1875 & 0.1642 & 0.0720 & 0.2061 
      & 0.2319 & \underline{0.2430} 
      & 0.2219 & 0.2327
      & 0.1478 & 0.2377 & 0.2435 & \textbf{0.2485} & 2.26$^*$\\
      %& 0.2410 & \textbf{0.2462} & 1.32$^*$\\
      & NDCG@20 & 0.1993 & 0.1760 & 0.0869 & 0.2178 
      & 0.2435 & \underline{0.2545} 
      & 0.2340 & 0.2454
      & 0.1600 & 0.2496 & 0.2558 & \textbf{0.2615} & 2.75$^*$\\
      %& 0.2533 & \textbf{0.2595} & 1.96$^*$\\
      \midrule
      \multirow{4}{*}{\MLOM}
      & Recall@10 & 0.1614 & \underline{0.2874} & 0.2349 & 0.2428 & 0.2745 & 0.2623 
      & 0.2397 & 0.2598
      & 0.1611 & 0.2409 
      & 0.2866 & \textbf{0.3000} & 4.38$^*$\\
      %& 0.2747 & \textbf{0.2916} & 1.46\textcolor{white}{$^*$}\\
      & Recall@20 & 0.2298 & \underline{0.3955} & 0.3422 & 0.3439 & 0.3707 & 0.3709 
      & 0.3455 & 0.3737
      & 0.2291 & 0.3571 & 0.3993 & \textbf{0.4132} & 4.48$^*$\\
      %& 0.3841 & \textbf{0.3944} & -0.28\textcolor{white}{$^*$}\\
      \cline{2-15}
      & NDCG@10 & 0.0841 & \underline{0.1619} & 0.1252 & 0.1374 & 0.1576 & 0.1463 
      & 0.1296 & 0.1349
      & 0.0832 & 0.1261 & 0.1582 & \textbf{0.1684} & 4.01$^*$\\
      %0.1481 & \textbf{0.1627} & 0.49\textcolor{white}{$^*$}\\
      & NDCG@20 & 0.1013 & \underline{0.1892} & 0.1524 & 0.1629 & 0.1818 & 0.1681 
      & 0.1561 & 0.1636
      & 0.1003 & 0.1553 & 0.1867 & \textbf{0.1970} & 4.12$^*$\\
      %& 0.1738 & \textbf{0.1887} & -0.26\textcolor{white}{$^*$}\\
      \midrule
      \multirow{4}{*}{\MLTM}
      & Recall@10 & 0.0992 & 0.1739 & OOM & 0.1588 
      & 0.1673 & \underline{0.1922} 
      & 0.1859 & 0.1915
      & 0.1106 & 0.1707 
      & \textbf{0.1957} & 0.1786 & 1.82$^*$\\
      %& \textbf{0.1924} & 0.1714 & 0.10\textcolor{white}{$^*$}\\ 
      & Recall@20 & 0.1622 & 0.2649 & OOM & 0.2390 
      & 0.2493 & \underline{0.2919} 
      & 0.2844 & 0.2909
      & 0.1866 & 0.2672 & \textbf{0.2959} & 0.2753 & 1.37$^*$\\
      %& \textbf{0.2924} & 0.2650 & 0.17\textcolor{white}{$^*$}\\
      \cline{2-15}
      & NDCG@10 & 0.0478 & 0.0907 & OOM & 0.0845 
      & 0.0895 & \underline{0.1004} 
      & 0.0967 & 0.1002
      & 0.0515 & 0.0862 & \textbf{0.1015} & 0.0924 & 1.10$^*$\\
      %& \textbf{0.0999} & 0.0879 & -0.50\textcolor{white}{$^*$}\\
      & NDCG@20 & 0.0636 & 0.1136 & OOM & 0.1047 
      & 0.1102 & \underline{0.1255} 
      & 0.1214 & 0.1251
      & 0.0706 & 0.1105 & \textbf{0.1267} & 0.1167 & 0.96$^*$\\
      %& \textbf{0.1251} & 0.1114 & -0.32\textcolor{white}{$^*$}\\
      \bottomrule
      \end{tabular}
      \begin{tablenotes}[normal,flushleft]
      \begin{footnotesize}
      \item
      For each dataset, the best performance in \methodmu and \method is in \textbf{bold}, 
      and the best performance among the baseline methods is \underline{underlined}.
      The column ``imprv" presents the percentage improvement of \method or \methodmu
      over the best-performing baseline methods. 
      %(\underline{underlined}).
      %
      The ${^*}$ indicates that the improvement is statistically significant at 95\% confidence level.
      \par
      \end{footnotesize}
      %\end{scriptsize}
      \end{tablenotes}
  \vspace{-10pt}
  \end{threeparttable}
\end{table*}

%% file: tables/summary.tex
\begin{table}
\footnotesize
%\vspace{-5pt}
  \caption{\mbox{{Performance Improvement of {\methodmu} and {\method} (\%)}}}
  \centering
  \label{tbl:summary}
  \vspace{-10pt}
  \begin{threeparttable}
      \begin{tabular}{
        @{\hspace{0pt}}l@{\hspace{16pt}}
        @{\hspace{16pt}}l@{\hspace{16pt}}
        @{\hspace{16pt}}r@{\hspace{16pt}}
        @{\hspace{16pt}}r@{\hspace{16pt}}
        @{\hspace{16pt}}r@{\hspace{0pt}}
        }
        \toprule
        Method & Metric & \Caser & \HAM & \SASRec\\
        \midrule
        \multirow{4}{*}{\methodmu}        
        & Recall@$10$  & 28.2$^*$ & 4.6$^*$ & 7.0$^*$\\
        & Recall@$20$  & 28.5$^*$ & 6.5$^*$ & 5.7$^*$\\
        & NDCG@$10$    & 29.6$^*$ & 2.5\textcolor{white}{$^*$} & 7.2$^*$\\
        & NDCG@$20$    & 29.8$^*$ & 3.8$^*$ & 7.0$^*$\\
        \midrule
        \multirow{4}{*}{\method}
        & Recall@$10$  & 32.8$^*$ & 7.8$^*$ & 10.6$^*$\\
        & Recall@$20$  & 32.5$^*$ & 9.3$^*$ &  8.7$^*$\\
        & NDCG@$10$    & 34.7$^*$ & 6.0$^*$ & 11.2$^*$\\
        & NDCG@$20$    & 34.4$^*$ & 7.0$^*$ & 10.6$^*$\\
        \bottomrule
      \end{tabular}
      \begin{tablenotes}[normal,flushleft]
      \begin{scriptsize}
      \item
      In this table, the column \Caser, \HAM and \SASRec represents the percentage improvement of \methodmu and \method over the corresponding method.  
      The $^*$ indicates that the improvement is statistically significant at 85\% confidence level.
    \par
      \end{scriptsize}
      \end{tablenotes}
  \end{threeparttable}
  \vspace{-10pt}
\end{table}

%% file: tables/notations.tex
\begin{table}[!h]
  \caption{Key Notations}
  \vspace{-10pt}
  \label{tbl:notations}
  \centering
  \begin{threeparttable}
     \begin{footnotesize}
      \begin{tabular}{
	@{\hspace{3pt}}l@{\hspace{3pt}}
	@{\hspace{3pt}}p{0.35\textwidth}@{\hspace{3pt}}          
	}
        \toprule
        notations & meanings \\
        \midrule
        $\mathbb{U}$    &  the set of users \\
        $\mathbb{V}$    &  the set of items \\
        $S_i$		&  the historical interaction sequence of user $u_i$ \\
        $v_{s_t}(i)$ & the $t$-th item in $S_i$ \\
        $B_i$    & the fixed-length sequence converted from $S_i$\\
        $v_{b_t}(i)$ & the $t$-th item in $B_i$ \\
        $v_g(i)$ & the ground-truth next item that user $u_i$ will interact with \\
        \bottomrule
      \end{tabular}
      \end{footnotesize}
  \end{threeparttable}
\end{table}

%% file: tables/parameters.tex
\begin{table*}
\footnotesize
  \caption{\mbox{Best-performing Hyper-parameters in \method, \methodmu and Baseline Methods}}
  \centering
  \label{tbl:para}
  %\vspace{-6pt}6pt
  \begin{threeparttable}
      \begin{tabular}{
        @{\hspace{4pt}}l@{\hspace{4pt}}
        @{\hspace{4pt}}r@{\hspace{4pt}}
        @{\hspace{6pt}}c@{\hspace{6pt}}
        @{\hspace{4pt}}r@{\hspace{4pt}}
        @{\hspace{4pt}}r@{\hspace{4pt}}
        @{\hspace{4pt}}r@{\hspace{4pt}}
        @{\hspace{4pt}}r@{\hspace{4pt}}
        @{\hspace{4pt}}r@{\hspace{4pt}}
        @{\hspace{6pt}}c@{\hspace{6pt}}
        @{\hspace{4pt}}r@{\hspace{4pt}}
        @{\hspace{4pt}}r@{\hspace{4pt}}
        @{\hspace{6pt}}c@{\hspace{6pt}}
        @{\hspace{4pt}}r@{\hspace{4pt}}
        @{\hspace{4pt}}r@{\hspace{4pt}}
        @{\hspace{4pt}}r@{\hspace{4pt}}
        @{\hspace{4pt}}r@{\hspace{4pt}}
        @{\hspace{6pt}}c@{\hspace{6pt}}
        @{\hspace{4pt}}r@{\hspace{4pt}}
        @{\hspace{4pt}}r@{\hspace{4pt}}
        @{\hspace{4pt}}r@{\hspace{4pt}}
        @{\hspace{4pt}}r@{\hspace{4pt}}
        @{\hspace{4pt}}r@{\hspace{4pt}}
        @{\hspace{4pt}}r@{\hspace{4pt}}
        @{\hspace{6pt}}c@{\hspace{6pt}}
        @{\hspace{4pt}}r@{\hspace{4pt}}
        @{\hspace{4pt}}r@{\hspace{4pt}}
        @{\hspace{4pt}}r@{\hspace{4pt}}
        @{\hspace{4pt}}r@{\hspace{4pt}}
        }
        \toprule
        \multirow{2}{*}{Dataset} 
        & \multicolumn{1}{c}{\FPMC}
        && \multicolumn{5}{c}{\Caser}  
        && \multicolumn{2}{c}{\NARM} 
        && \multicolumn{4}{c}{\HGN} 
        && \multicolumn{6}{c}{\HAM} 
        && \multicolumn{4}{c}{\SASRec}\\
        \cmidrule(lr){2-2}
        \cmidrule(lr){4-8} 
        \cmidrule(lr){10-11} 
        \cmidrule(lr){13-16} 
        \cmidrule(lr){18-23} 
        \cmidrule(lr){25-28}
        & $d$
	  && $d$ & $n_s$ & $n_p$ & $n_v$ & $n_f$
        && $d$ & $l_r$
        && $d$ & $n_s$ & $n_p$ & $\lambda$
        && $d$ & $n_s$ & $n_p$ & $\lambda$ & $n_l$ & $n_o$
        && $d$ & $n$ & $n_h$ & $n_b$ \\
        %&& $d$ & $n$ & $n_h$ & $n_b$
        %&& $d$ & $n$ & $n_h$ & $n_b$\\
        \midrule
        \Beauty & 512
        && 512 & 4 & 1 & 2 & 8  
        && 512 & 1e-3 
        && 512 & 3 & 2 & 1e-3 
        && 512 & 3 & 2 & 1e-3 & 1 & 1
        && 128 & 75  & 4 & 2 \\
        %&& 256 & 75  & 4 & 2 
        %&& 256 & 75 & 16 & 3\\ 
	  \Toys	& 512
        && 512 & 4 & 1 & 1 & 4  
        && 512 & 1e-3 
        && 512 & 3 & 1 & 1e-4 
        && 512 & 3 & 1 & 1e-3 & 1 & 1
        && 128 & 50  & 2 & 3 \\
        %&& 256 & 50  & 4 & 4 
        %&& 256 & 50 & 8 & 4\\
	  \Children & 128
        && 512 & 4 & 1 & 2 & 4  
        && 256 & 1e-4 
        && 256 & 3 & 1 & 0    
        && 256 & 4 & 2 & 1e-4 & 1 & 2
        && 256 & 175 & 2 & 1 \\
        %&& 256 & 200 & 1 & 2 
        %&& 512 & 100 & 1 & 1\\
	  \Comics & 256
        && 512 & 4 & 1 & 1 & 4  
        && 256 & 1e-3 
        && 512 & 3 & 1 & 0 
        && 512 & 3 & 1 & 0 & 1 & 3
        && 256 & 200 & 1 & 1 \\
        %&& 512 & 200 & 1 & 1 
        %&& 512 & 200 & 1 & 1\\
	  \MLOM & 256
        && 128 & 6 & 1 & 4 & 16  
        && 512 & 1e-4 
        && 128 & 4 & 1 & 1e-3 
        && 512 & 5 & 1 & 1e-3 & 2 & 1
        && 256 & 150 & 4 & 3 \\
        %&& 512 & 200 & 2 & 5 
        %&& 512 & 200 & 4 & 4\\
	  \MLTM & 128 
        && 256 & 6 & 1 & 2 & 8  
        && OOM & OOM  
        && 128 & 4 & 2 & 1e-3 
        && 256 & 5 & 2 & 1e-3 & 3 & 3
        && 256 & 150 & 2 & 3 \\
        %&& 512 & 150 & 8 & 5 
        %&& 256 & 100 & 4 & 4\\
        %\bottomrule
      \end{tabular}
      \begin{tabular}{
        @{\hspace{4pt}}l@{\hspace{4pt}}
        @{\hspace{4pt}}r@{\hspace{4pt}}
        @{\hspace{4pt}}r@{\hspace{4pt}}
        @{\hspace{4pt}}r@{\hspace{4pt}}
        @{\hspace{4pt}}r@{\hspace{4pt}}
        @{\hspace{6pt}}c@{\hspace{6pt}}
        @{\hspace{4pt}}r@{\hspace{4pt}}
        @{\hspace{4pt}}r@{\hspace{4pt}}
        @{\hspace{4pt}}r@{\hspace{4pt}}
        @{\hspace{4pt}}r@{\hspace{4pt}}
        @{\hspace{6pt}}c@{\hspace{6pt}}
        @{\hspace{4pt}}r@{\hspace{4pt}}
        @{\hspace{4pt}}r@{\hspace{4pt}}
        @{\hspace{4pt}}r@{\hspace{4pt}}
        @{\hspace{6pt}}c@{\hspace{6pt}}
        @{\hspace{4pt}}r@{\hspace{4pt}}
        @{\hspace{4pt}}r@{\hspace{4pt}}
        @{\hspace{4pt}}r@{\hspace{4pt}}
        @{\hspace{4pt}}r@{\hspace{4pt}}
        @{\hspace{6pt}}c@{\hspace{6pt}}
        @{\hspace{4pt}}r@{\hspace{4pt}}
        @{\hspace{4pt}}r@{\hspace{4pt}}
        @{\hspace{4pt}}r@{\hspace{4pt}}
        @{\hspace{4pt}}r@{\hspace{4pt}}
        @{\hspace{6pt}}c@{\hspace{6pt}}
        @{\hspace{4pt}}r@{\hspace{4pt}}
        @{\hspace{4pt}}r@{\hspace{4pt}}
        @{\hspace{4pt}}r@{\hspace{4pt}}
        @{\hspace{4pt}}r@{\hspace{4pt}}
        }
        \midrule
        \multirow{2}{*}{Dataset} 
        & \multicolumn{4}{c}{\SC}
        && \multicolumn{4}{c}{\SM}
        &&\multicolumn{3}{c}{\BERT}
        &&\multicolumn{4}{c}{\FMLP}
        &&\multicolumn{4}{c}{\methodmu} 
        &&\multicolumn{4}{c}{\method}\\
        \cmidrule(lr){2-5} 
        \cmidrule(lr){7-10} 
        \cmidrule(lr){12-14} 
        \cmidrule(lr){16-19} 
        \cmidrule(lr){21-24}
        \cmidrule(lr){26-29}
        &  $d$ & $n$ & $n_h$ & $n_b$
        && $d$ & $n$ & $n_h$ & $n_b$
        && $d$ & $n_h$ & $n_b$
        && $d$ & $n$ & $n_h$ & $n_b$
        && $d$ & $n$ & $n_h$ & $n_b$
        && $d$ & $n$ & $n_h$ & $n_b$\\
        \midrule
        \Beauty   
        &  256 & 50 & 4 & 3
        && 128 & 50 & 2 & 3
        && 512 & 4 & 1
        && 128 & 200 & 1 & 3
        && 256 & 75  & 4 & 2 
        && 256 & 75 & 16 & 3\\ 
	  \Toys	   
        &  256 & 75 & 8 & 5 
        && 128 & 100 & 16 & 3
        && 512 & 4 & 1
        && 128 & 50 & 1 & 3
        && 256 & 50  & 4 & 4 
        && 256 & 50 & 8 & 4\\
	  \Children  
        &  256 & 100 & 2 & 2
        && 256 & 200 & 2 & 2
        && 256 & 4 & 2
        && 256 & 150 & 1 & 1
        && 256 & 200 & 1 & 2 
        && 512 & 100 & 1 & 1\\
	  \Comics   
        &  256 & 200 & 1 & 2
        && 256 & 100 & 1 & 3
        && 128 & 4 & 1
        && 512 & 200 & 1 & 1
        && 512 & 200 & 1 & 1
        && 512 & 200 & 1 & 1\\
	  \MLOM     
        &  256 & 75 & 4 & 3
        && 512 & 200 & 2 & 4
        && 256 & 4 & 3
        && 512 & 175 & 1 & 1
        && 512 & 200 & 2 & 5 
        && 512 & 200 & 4 & 4\\
	  \MLTM     
        &  256 & 200 & 2 & 4
        && 512 & 150 & 2 & 2
        && 128 & 4 & 3
        && 512 & 150 & 1 & 2
        && 512 & 150 & 8 & 5 
        && 256 & 100 & 4 & 4\\
        \bottomrule
      \end{tabular}
      \begin{tablenotes}[normal,flushleft]
      \begin{scriptsize}
      \item
      In the table. ``OOM" represents the out-of-memory issue.
      \par
      \end{scriptsize}
      \end{tablenotes}
      \vspace{-10pt}
  \end{threeparttable}
\end{table*}

%% file: tables/dataset.tex
\begin{table}
\footnotesize
  \caption{Dataset Statistics}
  \centering
  %\vspace{-8pt}
  \label{tbl:dataset}
  \begin{threeparttable}
      \begin{tabular}{
	@{\hspace{9pt}}l@{\hspace{9pt}}
	@{\hspace{9pt}}r@{\hspace{9pt}}      
	@{\hspace{9pt}}r@{\hspace{9pt}}
	@{\hspace{9pt}}r@{\hspace{9pt}}
	@{\hspace{9pt}}r@{\hspace{9pt}}
    @{\hspace{9pt}}r@{\hspace{9pt}}
	}
        \toprule
        dataset      & \#users   & \#items &  \#intrns & {\#intrns/u} & \#u/i\\
        \midrule
        \Beauty       & 22,363  & 12,101  & 198,502    &  8.9  & 16.4 \\
        \Toys         & 19,412  & 11,924  & 167,597    &  8.6  & 14.1 \\
        \Children     & 48,296  & 32,871  & 2,784,423  & 57.6  & 84.7 \\
        \Comics       & 34,445  & 33,121  & 2,411,314  & 70.0  & 72.8 \\
        \MLOM         & 6,040   & 3,952   & 1,000,209  & 165.6 & 253.1\\
        \MLTM         & 129,780 & 13,663  & 9,926,480  & 76.5  & 726.5\\
        \bottomrule
      \end{tabular}
      \begin{tablenotes}[normal,flushleft]
      \begin{footnotesize}
      \item 
      In this table, ``\#users", ``\#items" and ``\#intrns" represents the number of 
      users, items and user-item interactions, respectively. 
      The column ``\#intrns/u" has the average 
      number of interactions of each user. 
      The column ``\#u/i" has the average number of interactions on each item.  
      \par
      \end{footnotesize}
      \end{tablenotes}
  \end{threeparttable}
  \vspace{-8pt}
\end{table}